\documentclass{vldb}

\usepackage{graphicx}
\usepackage{balance}  

\usepackage[colorlinks=true,linkcolor=blue,citecolor=blue,urlcolor=blue]{hyperref}
\usepackage[lined,boxed,linesnumbered,commentsnumbered]{algorithm2e}

\usepackage{etoolbox}
\usepackage{comment}

\def\equationautorefname~#1\null{%
  (#1)\null
}

\usepackage{subcaption}
\usepackage[group-separator={,},group-minimum-digits=4]{siunitx}
\usepackage{multirow}
\usepackage{bbm}
\usepackage{diagbox}
\usepackage[table]{xcolor}
\usepackage{array}
\usepackage{colortbl}
\usepackage{mathtools}
\usepackage{makecell}

\colorlet{red}{black}

\usepackage{cleveref}
\Crefrangeformat{figure}{Figures.~(#3#1#4) to~(#5#2#6)}
\usepackage{chngcntr}

\usepackage{tikz-cd}

\newcommand{\RN}[1]{%
  \textup{\uppercase\expandafter{\romannumeral#1}}%
}
\usetikzlibrary{decorations.pathmorphing}

\newcommand\xrsquigarrow[1]{%
    \mathrel{%
        \begin{tikzpicture}[%
            baseline={(current bounding box.south)}
            ]
        \node[%
            ,inner sep=.44ex
            ,align=center
            ] (tmp) {$\scriptstyle #1$};
        \path[%
            ,draw,<-
            ,decorate,decoration={%
                ,zigzag
                ,amplitude=0.7pt
                ,segment length=1.2mm,pre length=3.5pt
                }
            ] 
        (tmp.south east) -- (tmp.south west);
        \end{tikzpicture}
        }
    }

\newcommand{\RNum}[1]{\uppercase\expandafter{\romannumeral #1\relax}}
\newcommand{\Rnum}[1]{\lowercase\expandafter{\romannumeral #1\relax}}
\def\code{\texttt}
\newenvironment{sitemize}{%
  \begin{list}{$\bullet$}{%
    \setlength{\itemsep}{0.1cm}%
    \setlength{\leftmargin}{1.0em}%
    \setlength{\topsep}{0.1cm}%
    \setlength{\parsep}{0mm}}%
  }{\end{list}}

\usepackage{etoolbox}

\usepackage{amsthm}
\newtheoremstyle{saber}
  {}
  {}
  {\itshape}
  {}
  {\bfseries}
  {.}
  {.5em}
  {}

\theoremstyle{saber}
\ifcsundef{theorem}{
\newtheorem{theorem}{Theorem}
}{}
\ifcsundef{corollary}{%
\newtheorem{corollary}{Corollary}
}{}
\ifcsundef{lemma}{%
}{}
\ifcsundef{definition}{%
}{}
\ifcsundef{fact}{%
\newtheorem{fact}{Fact}
}{}
\ifcsundef{remark}{%
}{}
\ifcsundef{property}{%
}{}

\ifx\theoremautorefname\undefined
\newcommand*{\theoremautorefname}{Theorem}
\fi
\ifx\lemmaautorefname\undefined
\newcommand*{\lemmaautorefname}{Lemma}
\fi
\ifx\definitionautorefname\undefined
\newcommand*{\definitionautorefname}{Definition}
\fi
\ifx\corollaryautorefname\undefined
\newcommand*{\corollaryautorefname}{Corollary}
\fi
\ifx\factautorefname\undefined
\newcommand*{\factautorefname}{Fact}
\fi
\ifx\propertyautorefname\undefined
\newcommand*{\propertyautorefname}{Property}
\fi

\ifx\argmax\undefined
\newcommand{\argmax}{\operatornamewithlimits{argmax}}
\fi
\ifx\argmin\undefined
\newcommand{\argmin}{\operatornamewithlimits{argmin}}
\fi
\ifx\limsup\undefined
\newcommand{\limsup}{\operatornamewithlimits{limsup}}
\fi
\ifx\liminf\undefined
\newcommand{\liminf}{\operatornamewithlimits{liminf}}
\fi
\ifx\norm\undefined
\newcommand{\norm}[1]{\lVert#1\rVert}
\fi
\ifx\abs\undefined
\newcommand{\abs}[1]{\lvert#1\rvert}
\fi
\ifx\set\undefined
\newcommand{\set}[1]{\left\{#1\right\}}
\fi
\ifx\mset\undefined
\newcommand{\mset}[1]{\lbrack #1\rbrack}
\fi
\ifx\etal\undefined
\newcommand{\etal}[1]{{\em #1 et al.}~}
\fi
\ifx\ie\undefined
\newcommand{\ie}{{\em i.e.,} }
\fi
\ifx\eg\undefined
\newcommand{\eg}{{\em e.g.,} }
\fi
\ifx\etc\undefined
\newcommand{\etc}{{\em etc.,} }
\fi
\ifx\wrt\undefined
\newcommand{\wrt}{{\em w.r.t.} }
\fi
\ifx\dotprod\undefined
\newcommand{\dotprod}[2]{
  \langle #1, #2 \rangle
}
\fi
\ifx\iid\undefined 
\newcommand{\iid}{i.i.d.}
\fi
\ifx\bigParenthes\undefined 
\newcommand{\bigParenthes}[1]{
  \big(#1\big)
}
\fi
\ifx\bigBracket\undefined 
\newcommand{\bigBracket}[1]{
  \big\{#1\big\}
}
\fi
\ifx\bigSqBracket\undefined 
\newcommand{\bigSqBracket}[1]{
  \big[#1\big]
}
\fi
\ifx\BigParenthes\undefined 
\newcommand{\BigParenthes}[1]{
  \Big(#1\Big)
}
\fi
\ifx\BigBracket\undefined 
\newcommand{\BigBracket}[1]{
  \Big\{#1\Big\}
}
\fi
\ifx\BigSqBracket\undefined 
\newcommand{\BigSqBracket}[1]{
  \Big[#1\Big]
}
\fi
\ifx\biggParenthes\undefined 
\newcommand{\biggParenthes}[1]{
  \bigg(#1\bigg)
}
\fi
\ifx\biggBracket\undefined 
\newcommand{\biggBracket}[1]{
  \bigg\{#1\bigg\}
}
\fi
\ifx\biggSqBracket\undefined 
\newcommand{\biggSqBracket}[1]{
  \bigg[#1\bigg]
}
\fi
\ifx\BiggParenthes\undefined 
\newcommand{\BiggParenthes}[1]{
  \Bigg(#1\Bigg)
}
\fi
\ifx\BiggBracket\undefined 
\newcommand{\BiggBracket}[1]{
  \Bigg\{#1\Bigg\}
}
\fi
\ifx\BiggSqBracket\undefined 
\newcommand{\BiggSqBracket}[1]{
  \Bigg[#1\Bigg]
}
\fi
\ifx\bracket\undefined
\newcommand{\bracket}[1]{
  \{#1\}
}
\fi
\ifx\parenthes\undefined
\newcommand{\parenthes}[1]{
  (#1)
}
\fi
\ifx\sqBracket\undefined
\newcommand{\sqBracket}[1]{
  [#1]
}
\fi
\ifx\prob\undefined 
\newcommand{\prob}[1]{\mathbb{P}[#1]}
\fi
\ifx\Prob\undefined 
\newcommand{\Prob}[1]{\mathbb{P}\big[#1\big]}
\fi
\ifx\probb\undefined 

\fi
\ifx\expect\undefined 
\newcommand{\expect}[1]{\mathbb{E}[#1]}
\fi
\ifx\Expect\undefined 
\newcommand{\Expect}[1]{\mathbb{E}\big[#1\big]}
\fi
\ifx\expectt\undefined 
\newcommand{\expectt}[1]{\mathbb{E}\bigg[#1\bigg]}
\fi
\ifx\walk\undefined 
\makeatletter
\newcommand{\walk}[1]{%
  \@tempswafalse
  \@for\next:=#1\do
    {\if@tempswa\!\!\rightarrow\!\!\else\@tempswatrue\fi\next}%
}
\makeatother
\fi
\ifx\seq\undefined 
\newcommand{\seq}{\!=\!}
\fi
\ifx\sminus\undefined 
\newcommand{\sminus}{\!-\!}
\fi
\ifx\sm\undefined 
\newcommand{\sm}[1]{\!#1\!}
\fi

\ifx\union\undefined
\newcommand{\union}[2]{#1\!\cup\!#2}
\fi

\usepackage{breakurl}
\newcommand\refToAppendix[2]{#1}

\vldbTitle{Space- and Computationally-Efficient Set Reconciliation via Parity Bitmap Sketch (PBS)}
\vldbAuthors{Long Gong, Ziheng Liu, Liang Liu, Jun Xu, Mitsunori Ogihara, Tong Yang}
\vldbDOI{https://doi.org/10.14778/xxxxxxx.xxxxxxx}
\vldbVolume{14}
\vldbNumber{xxx}
\vldbYear{2021}

\begin{document}
\sloppy


\title{Space- and Computationally-Efficient Set Reconciliation via Parity Bitmap Sketch (PBS)}

\author{
\end{tabular}%
\begin{tabular}{cccccc}
Long Gong$^\dagger$
&
Ziheng Liu$^{\dagger\dagger}$
&
Liang Liu$^\dagger$
&
Jun Xu$^\dagger$
&
Mitsunori Ogihara$^\ddagger$
&
Tong Yang$^{\dagger\dagger}$
\end{tabular}%
\\
\begin{tabular}{ccccc}
$^\dagger$\affaddr{Georgia Institute of Technology, USA} && $^{\dagger\dagger}$\affaddr{Peking University, China} && $^\ddagger$\affaddr{University of Miami, USA}\\
{\small \texttt{$\{$gonglong,lliu315$\}$@gatech.edu}  \texttt{jx@cc.gatech.edu}} && {\small \texttt{$\{$liuziheng,yang.tong$\}$@pku.edu.cn}}&&{\small \texttt{ogihara@cs.miami.edu} } 
}

\maketitle

\begin{abstract}
Set reconciliation is a fundamental algorithmic problem that arises in many networking, system, and database applications. 
In this problem, two large sets $A$ and $B$ of objects (bitcoins, files, records, etc.)
are stored respectively at two different network-connected hosts, which we name Alice and Bob respectively.
Alice and Bob communicate with each other to learn $A\triangle B$, the 
difference between $A$ and $B$, and as a result the reconciled set $A\bigcup B$. 

Current set reconciliation schemes are based on either invertible Bloom filters (IBF) or error-correction codes (ECC). The former has a low computational complexity of $O(d)$, where $d$ is the 
cardinality of $A\triangle B$, but has a high communication overhead that is several times larger than the theoretical minimum. 
The latter has a low communication overhead close to the theoretical minimum, but has a much higher computational complexity of $O(d^2)$. 
In this work, we propose Parity Bitmap Sketch (PBS), an ECC-based set reconciliation scheme that gets the better of both worlds: 
PBS has both a low computational complexity of $O(d)$ just like IBF-based solutions and a low communication overhead of roughly twice the theoretical minimum. 
A separate contribution of this work is a novel rigorous analytical framework that can be used for the precise calculation of various performance metrics and for the near-optimal parameter tuning of PBS.
\end{abstract}

\section{Introduction}\label{sec:intro}

Set reconciliation is a fundamental algorithmic problem that has received considerable research attention over the past two 
decades~\cite{Minsky2003,Guo_SetReconciliationvia_2013,Luo_SetReconciliationCuckoo_2019,Eppstein_WhatsDifference?Efficient_2011,Dodis_PinSketch_2008}. 
In the simplest form of this problem, two large sets $A$ and $B$ of objects (bitcoins, files, records, etc.)
are stored respectively at two different network-connected hosts, which we name Alice and Bob respectively.
Alice and Bob communicate with each other to find out the 
difference between $A$ and $B$, defined as $A\triangle B\triangleq (A\! \setminus\! B) \bigcup (B\! \setminus\! A)$,
so that both Alice and Bob obtain the set union $A\bigcup B$ $(=A\bigcup (A\triangle B)=B\bigcup (A\triangle B))$.  

Set reconciliation arises in many networking, system, and database applications.  
In cloud storage systems ({\it e.g.,} Dropbox, Microsoft OneDrive, Google Drive, and Apple iCloud), sets of files and directories 
need to be synchronized across the copies stored locally on different devices and in the cloud. 
In distributed database systems ({\it e.g.,} Spanner~\cite{Corbett_Spanner_2013} and Cassandra~\cite{Lakshman_Cassandra_2010}), 
an update at a single node has to get replicated across all other nodes eventually. 
In blockchains~\cite{Ozisik2019,Naumenko_ErlayEfficientTransaction_2019}, transactions need to be synchronized with some peers.

\subsection{Problem Formulation}\label{sec:intro_prob_formulation}

As is standard in the literature, in the rest of the paper we describe only unidirectional set reconciliation, in which Alice learns $A\triangle B$ and then infers $A\bigcup B$;
for bidirectional set reconciliation, Alice can simply infer $A\!\setminus\! B$ (from $A\triangle B$) and send it to Bob, from which Bob can infer $A\bigcup B$ as well. 
A simple but naive set reconciliation scheme is for Bob to send $B$, in its entirety, to Alice.  
This scheme, however, is grossly inefficient when
$A\triangle B$ is small (in cardinality) relative to their union $A \bigcup B$,
which is indeed a usual situation in most applications. 
In this situation, it would be ideal if only the elements (objects) in $A\triangle B$ need to be transmitted.  In other words, 
Bob sends only $B\! \setminus\! A$ to Alice.



In the set reconciliation problem, we usually assume each element is ``indexed" by a fixed-length (hash) signature, so the universe $\mathcal{U}$ contains all binary strings of this length.
For example, when this length is $32$ bits, $\mathcal{U}$ contains
$2^{32}$ elements.  We denote this length as $\log |\mathcal{U}|$ in the sequel. 
Let $d\triangleq |A\triangle B|$ denote the cardinality 
of the set difference.   It is not hard to prove (using information theory) that the theoretical minimum amount of communication between Alice and Bob for the {\it bidirectional} 
set reconciliation is the size of the set difference $d \log |\mathcal{U}|$~\cite{Minsky2003}. 
{\color{red}
It is reasonable to use this minimum as a comparison benchmark for communication overheads  
in the unidirectional case (wherein Alice learns $A\triangle B$), because it is provably also the minimum for this unidirectional case 
in certain worst-case scenarios such as $A\subset B$.  
Hence we will do so throughout this paper.
}

\subsection{Existing Approaches}\label{sec:psea}

Although many techniques have been proposed for this problem, they all fall victim to a seemingly fundamental 
tradeoff between the communication overhead, of transmitting the codewords (in the general sense rather than in the narrow context of error-correction codes) needed for 
set reconciliation, and the computational complexity of decoding such codewords.  
The majority of such techniques are based on either invertible Bloom filters (IBF) or error-correction codes (ECC).  
On one hand, IBF-based techniques incur a communication overhead that is several times  
(e.g., $6$ times in~\cite{Eppstein_WhatsDifference?Efficient_2011}) the theoretical minimum $d \log |\mathcal{U}|$,
but have a linear (i.e., $O(d)$) decoding computational complexity.  
On the other hand, ECC-based techniques have a low communication overhead close to the theoretical minimum, but 
have a decoding computational complexity of $O(d^2)$ finite field operations,
which can be very high when $d$ is large (say $d \!=\! \num{10000}$).

\subsection{Our Solution}\label{sec:intro_oursolution}

In this work, we propose a solution, called Parity Bitmap Sketch (PBS), that mostly avoids this unfortunate tradeoff and gets the better of both worlds.  More specifically, PBS has both a low computational complexity of $O(d)$ just like IBF-based solutions and a low communication overhead of roughly twice the theoretical minimum.
PBS also has another advantage over all existing solutions in that it is ``piecewise reconciliable" in the following sense.  
In all existing solutions, decoding of the codewords to obtain $A\triangle B$ is an all-or-nothing process
in the sense that when the decoding failed 
(albeit usually with a small probability when the codewords are appropriately parameterized), 
little knowledge has been learned (so most of the communication, encoding, and decoding efforts are wasted) and the process starts from square one.  
In contrast, in PBS, the decoding of each codeword (also in the general sense) 
is independent of those of others, and the successful decoding of each codeword leads to a subset of 
{\it distinct elements} being reconciled;   here and in the sequel, we refer to each element in $A\triangle B$ a {\it distinct element}.  
This way,
additional efforts are incurred only for the small percentage of codewords whose decodings failed earlier.




The only minor tradeoff of our solution is that the number of rounds of message exchanges needed for set reconciliation is slightly larger 
than some existing solutions.  However, in almost all practical application scenarios, this tradeoff is not expected to lead to longer response time 
for the following reason.  
Thanks to the piecewise reconciliability of PBS, the vast majority, if not all, of the distinct elements in $A\triangle B$ are successfully reconciled in the first round.  
The ``synchronization'' of the objects ``indexed" by these successfully reconciled distinct elements can then run in parallel with the reconciliation of the rest in $A\triangle B$.  
In other words, the piecewise reconciliability is expected to effectively mask the slightly larger number of communication rounds as far as the response time is concerned.

\subsubsection{When \texorpdfstring{\larger\MakeLowercase{$d$}}{d} is small}\label{sec:smalld}


Here we sketch the main ideas of PBS.
{\color{red}For the moment, we assume that the set difference cardinality $d$ is small (say no more than $5$ elements), 
and the value of $d$ is precisely known.}    
The first step of PBS is to partition $A$ and $B$ each into subsets in a consistent manner.
We partition the set $A$ into $n$ disjoint subsets $A_1$, $A_2$, ..., $A_n$ using a hash function $h$ as follows:
$A_1$ contains all elements in $A$ that are hashed to value $1$ (by $h$), $A_2$ contains all elements in $A$ that are hashed to value $2$, and so on.  
In the same manner, we partition $B$ into $B_1, B_2, \cdots, B_n$ using the same $h$. 
The use of the common hash function $h$ induces a hash-partitioning (also by $h$)
of the set-pair-difference $A\triangle B$ into $n$ disjoint subset-pair-differences $A_1 \triangle B_1$, $A_2 \triangle B_2$, ..., 
$A_n \triangle B_n$. We set this constant $n$ to be roughly an order of magnitude larger than $d^2$, so that  
with high probability, the following {\it ideal situation} happens:
The $d$ distinct elements between $A$ and $B$ are hashed (by $h$) into $d$ distinct subset-pair-differences,
so that each such subset-pair-difference contains exactly one distinct element.  This situation is ideal, because 
each such subset pair 
can be easily reconciled as will be shown in \autoref{subsec:protocol0}. 
To guarantee that the ideal situation happens with high probability, the value of $n$ does not have to be very large when $d$ is small. 
For example, when $d = 5$ and $n$ is set to $255$, the probability for the ideal situation to occur is $0.96$.


The second step of PBS is to encode partitions $\{A_i\}_{i=1}^n$ and $\{B_i\}_{i=1}^n$ each into an $n$-bit-long {\it parity bitmap}.  
The $n$-bit-long parity bitmap encoding of $\{A_i\}_{i=1}^n$, denoted as $A[1..n]$, is defined as follows. 
For $i = 1, 2, ..., n$, $A[i]$, the $i^{th}$ bit of $A[1..n]$, is equal to $1$ if $A_i$ contains an odd number of elements,
and is equal to $0$ otherwise. 
The $n$-bit-long parity bitmap of $\{B_i\}_{i=1}^n$, denoted as $B[1..n]$, is similarly defined. 
In the aforementioned ideal situation of the $d$ elements in $A\triangle B$ landing in $d$ distinct subset-pair-differences, the two bitmaps differ in exactly $d$ bit positions.  In this situation, if Bob knows these $d$ bit positions then the $d$ corresponding subset pairs, and hence the set pair $A$ and $B$, can be easily reconciled as we will describe in \autoref{sec:pbs_smalld}.  

While Alice can certainly send the $n$-bit-long parity bitmap $A[1..n]$ to Bob, this is quite wasteful since $n\! \gg\! d$.  
{\color{red}A more communication-efficient way, introduced first in PinSkech~\cite{Dodis_PinSketch_2008}, is to 
view $B[1..n]$ as a ``corrupted" copy of $A[1..n]$ that contains $d$ ``bit errors" at the $d$ bit positions where $A[1..n]$ and $B[1..n]$ differ,
and to let Alice send Bob instead a BCH~\cite{bch1} codeword of much shorter length
that can ``correct'' $B[1..n]$ (which Bob already has locally) into $A[1..n]$.}
Referring to this BCH coding as sketching (as was done in~\cite{Dodis_PinSketch_2008}), we call our scheme Parity Bitmap Sketch (PBS).

\subsubsection{When \texorpdfstring{\larger\MakeLowercase{$d$}}{d} is large}\label{sec:larged}

{\color{red}In general, $d$ can be much larger than several (say $5$).} 
When $d$ is very large, it would be computationally too costly to decode all $d$ errors 
``in one shot'' (i.e., in a single parity bitmap) using the
aforementioned $O(d^2)$ BCH decoding algorithm~\cite{Wuille_Minisketchoptimizedlibrary_2019}.
{\color{red}Instead, in this case we partition, consistently using a different hash function $h'$ (than the $h$ above), 
sets $A$ and $B$ each into $d/\delta$ smaller sets, where 
$\delta$ is a small number 
(just like what we earlier assumed $d$ to be).}  
We refer to these smaller sets as {\it groups} to distinguish them from the {\it subsets} $A_i$'s and $B_i$'s above.  
With this partitioning, on average only $\delta$ distinct elements are hashed to any group pair.
Then, the PBS-for-small-$d$ scheme described above is used to reconcile each group pair. 
The computational complexity of the BCH decoding involved in reconciling each group pair is only $O(\delta^2)$, which can be considered $O(1)$, since $\delta$ is a small constant. 
As a result, the total BCH and other decoding computational complexity of PBS (for all $d/\delta$ group pairs) is only $O(d)$.

\subsubsection{Markov-chain modeling of PBS}

Another significant contribution of this work is a rigorous and accurate Markov-chain modeling of the multi-round set reconciliation process of PBS.  This model enables not only the
accurate analysis of various performance metrics, such as the probability that all distinct elements are successfully reconciled in $r$ rounds, but also the
tuning of the parameters of PBS for near-optimal performances.  In contrast, most existing solutions lack such a rigorous analytical framework. 

\subsubsection{Possible Applications of PBS}\label{subsec:db-app}

{\color{red}
As explained earlier, elements in set difference $A \triangle B$ are the hash signatures of actual objects that need to be exchanged.  
When the size of an object is much larger than that of a hash signature,  
the communication overhead of reconciling $A$ and $B$, using any existing set conciliation scheme except the naive scheme, 
is anyway negligible compared to that of 
exchanging the actual objects.   However, in many real-life applications, either
the actual object size is not significantly larger (e.g., in the transaction relay operation of a blockchain scheme), or the actual objects need to be 
synchronized mush less often than their hash signatures (e.g., in Dropbox under the smart sync mode~\cite{_dropbox_smart_sync_}).  In such applications, it makes a performance difference to 
reduce the communication overhead of reconciling $A$ and $B$.

For example, as measured in a blockchain work called Erlay~\cite{Naumenko_ErlayEfficientTransaction_2019}, 
this communication overhead accounts for around $5\%$ of the total network bandwidth consumption of its transaction relay operation.
In this case, 
PinSkech~\cite{Dodis_PinSketch_2008}, the most communication-efficient set reconciliation scheme, is used and 
the size of the hash signature (called {\it transaction ID} in blockchain schemes) is compressed from \SI{256}{bits} to \SI{64}{bits} 
(at cost of possible hash collisions among different transactions during the set reconciliation process).   This communication overhead would
increase to around $55\%$ of the total bandwidth consumption if IBF-based schemes were used instead and transaction ID's were not compressed.  

PBS is better suited, than any existing set reconciliation scheme, for such applications in general and blockchain schemes in particular, for two reasons.
First, although PinSketch, the state-of-the-art ECC-based solution, has a slightly smaller communication overhead than PBS, its  
computational complexity 
is too high to scale to scenarios  
where $| A\triangle B  |$ is large. 
Second, the communication overhead of IBF-based solutions, including the state-of-the-art solution Graphene~\cite{Ozisik2019}, are generally much larger than that of PBS.

Therefore, in the following we will use the transaction relay operation in blockchain schemes as an example application for PBS.
Transaction relay (operation) refers to the synchronization (reconciliation) of the transaction databases (sets) across 
the peer-to-peer network of a blockchain scheme.  In this application, Alice and Bob are two peers engaging in a transaction relay,
and $A$ and $B$ are the sets of hash signatures of the transactions recorded at Alice and Bob respectively. 
The blockchain schemes and their transaction relay operations have recently 
received considerable research attention from the database community ({\it e.g.,} ProvenDB~\cite{_ProvenDB_}, 
BlockchainDB~\cite{ElHindi_BlockchainDBSharedDatabase_2019}, 
BigchainDB~\cite{McConaghy_Bigchaindbscalableblockchain_2016}, 
and SEBDB~\cite{Zhu_SEBDBSemanticsEmpowered_2019}), partly due to the semantic similarities between blockchains and 
distributed databases~\cite{Ruan_BlockchainsDistributedDatabases_2019,Sharma_BlurringLinesBlockchains_2019}.
}

The rest of the paper is organized as follows.  First, we describe the aforementioned PBS-for-small-$d$ and PBS-for-large-$d$ schemes in 
\autoref{sec:scheme} and~\autoref{sec:pbs-for-large-d}, respectively.
Then, we describe our analytical framework in \autoref{sec:tail_bound} and apply it to the performance analysis and the near-optimal parameter tuning of PBS in \autoref{sec:application_analysis}. After that, we present a new estimator for estimating the set difference cardinality in~\autoref{subsec:ToW}. 
Finally, we survey existing set reconciliation schemes in \autoref{sec:related}, 
compare the performance of PBS with that of some of them in \autoref{sec:evaluation}, and conclude the paper in \autoref{sec:conclusion}.

\section[PBS for small \texorpdfstring{$d$}{d}]{PBS for small \texorpdfstring{\larger\boldmath\MakeLowercase{$d$}}{d}}\label{sec:scheme} 

In this section, we describe how the PBS scheme allows Alice and Bob to reconcile their respective sets $A$ and $B$, where $d\!=\!|A\triangle B|$ is assumed to be small and precisely known. 
We start with the trivial case where $d\!\leq\! 1$ in \autoref{subsec:protocol0} and then generalize the scheme for the case where $d$ is a small number in \autoref{sec:pbs_smalld}. 
{\color{red}As will be explained later in~\autoref{sec:BCH_code}, the latter will use the former as a building block.}

\subsection{The Trivial Case of \texorpdfstring{\larger\boldmath $d\leqslant 1$}{d<2}}\label{subsec:protocol0}

\begin{algorithm}
\caption{Set reconciliation when $d\le 1$}
\label{alg:trivial_case}

{\it Bob}: $s_B\leftarrow\underset{b\in B}{\oplus} b$; Send $s_B$ to Alice\; \label{alg1:sender}

{\it Alice}: $s_A\leftarrow\underset{a\in A}{\oplus} a$; $s\leftarrow s_A \oplus s_B$\; \label{alg1:receiver}
\end{algorithm}

\autoref{alg:trivial_case} shows the set reconciliation scheme for the trivial case, in which $A$ and $B$ differ by at most one (distinct) element. 
It consists of two steps. First, Bob calculates the \code{XOR} sum $s_B$, the bitwise-\code{XOR} of all elements in $B$, and sends it to Alice. 
Second, Alice calculates $s_A$, the \code{XOR} sum of all elements in $A$.  Upon receiving $s_B$ from Bob, Alice computes $s\!\triangleq\! s_A \!\oplus\! s_B$. 
The value of $s$ tells Alice which of the following two cases happens.

\begin{sitemize}
	\item Case (\RN{1}): If $s=\mathbf{0}$, which implies $s_A=s_B$, Alice concludes that $A$ and $B$ have no distinct element or $A=B$. 
	Here $\mathbf{0}$ denotes the $\log |\mathcal{U}|$-bit-long all-$0$ string;
	\item Case (\RN{2}): If $s\not=\mathbf{0}$, which implies $s_A\not=s_B$, Alice concludes that $A$ and $B$ have exactly one distinct element which is $s$, i.e., 
	$A\triangle B=\{s\}$.  This is because \code{XOR}ing $s_A$ and $s_B$ (to obtain $s$) cancels out all (common) elements in $A\bigcap B$.   
\end{sitemize}

Like in most of the literature on set reconciliation, we assume that the all-$0$ element 
$\mathbf{0}$ is excluded from the universe  $\mathcal{U}$, since otherwise \autoref{alg:trivial_case} does not work 
for the following reason.  When the computed $s$ is $\mathbf{0}$, Alice cannot tell whether $A$ and $B$ are identical, or they have $\mathbf{0}$ as their distinct element. 

\autoref{alg:trivial_case} also does not work when there are more than one distinct elements in $A\triangle B$ (i.e., $d>1$), since the computed $s$ in this case 
is the \code{XOR} sum of all these distinct elements.




\subsection{The General Case}\label{sec:pbs_smalld}

{\color{red}In this section, we describe the scheme for the more general case where $d$ is 
a small number (say $5$), 
but is not necessarily $0$ or $1$.}

\subsubsection{Hash-partitioning and parity bitmap encoding}\label{sec:hash_partition}

Here we formalize the aforementioned process of partitioning $A$ into $\{A_i\}_{i=1}^n$, $B$ into $\{B_i\}_{i=1}^n$, and $A\triangle B$ into $\{A_i\triangle B_i\}_{i=1}^n$ 
using a hash function $h$. 
Define {\it sub-universe} $\mathcal{U}_i$ as the set of elements in the universe $\mathcal{U}$ that are hashed into value $i$. More precisely, $\mathcal{U}_i\triangleq\{s\!\mid\! s\in \mathcal{U} \mbox{ and } h(s)=i\}$ for $i=1,2,...,n$. Then defining $A_i\triangleq A\bigcap\mathcal{U}_i$ and $B_i\triangleq B\bigcap\mathcal{U}_i$ for $i=1,2,...,n$ induces the partitioning of $A$, $B$, and $A\triangle B$.

How the $d$ distinct elements (balls) in $A\triangle B$ are ``scattered" into the $n$ subset-pair-differences (bins) $\{A_i\triangle B_i\}_{i=1}^n$ can be precisely modeled as throwing $d$ balls each uniformly and randomly into one of the $n$ bins. 
For the moment, we assume the following {\it ideal case} happens: Every subset-pair-difference $A_i\triangle B_i$ contains at most one distinct element.  
This ideal case corresponds to the $d$ balls ending up in $d$ distinct bins. It happens with probability $\prod_{k=1}^{d - 1}(1 - \frac{k}{n})$, which is on the order of $1-O(d^2/n)$ when $n\!\gg\!d$.
Hence, $n$ must be $\Omega(d^2)$ to ensure the ideal case happens with a nontrivial probability, as mentioned earlier in \autoref{sec:smalld}. 
\subsubsection{Find and reconcile the \texorpdfstring{$d$}{d} subset pairs}\label{sec:BCH_code}

\begin{algorithm}
\caption{PBS-for-small-$d$ (first round)}
\label{alg:single_round_pbs}


{\it Alice}: Send $\xi_A$, the BCH codeword of $A[1..n]$, to Bob\; \label{step:BCH_encode}

{\it Bob}: {\color{red}Decode $B[1..n]\|\xi_A$ to obtain $i_1, i_2, ..., i_d$\;} \label{step:BCH_decode}

{\it Bob}: Send \code{XOR} sums of sets $B_{i_1}, B_{i_2},...,B_{i_d}$ (\autoref{alg1:sender} in \autoref{alg:trivial_case}), indices $i_1, i_2, ..., i_d$,
and checksum $c(B)$ to Alice\; \label{step:send_xor}

{\it Alice}: Obtain $A_{i_1}\triangle B_{i_1}, A_{i_2}\triangle B_{i_2},...,$ and $A_{i_d}\triangle B_{i_d}$ (\autoref{alg1:receiver} in \autoref{alg:trivial_case}); $\hat{D}\leftarrow \bigcup_{k=1}^d (A_{i_k}\triangle B_{i_k})$\; \label{step:trivial_reconcile}

{\it Alice}: Check whether $c(A\triangle\hat{D})\stackrel{?}{=}c(B)$\; \label{step:terminate_condition}

\end{algorithm}

The remaining steps of the PBS scheme are summarized in \autoref{alg:single_round_pbs}.
Recall that the partitions $\{A_i\}_{i=1}^n$ and $\{B_i\}_{i=1}^n$ can be encoded as parity bitmaps $A[1..n]$ and $B[1..n]$ respectively, in which each $A[i]$ or $B[i]$
corresponds to the parity (oddness or evenness) of the cardinality of the subset $A_i$ or $B_i$.  In the ideal case, $A[1..n]$ and $B[1..n]$ differ in exactly $d$ distinct bit positions. 
Suppose these $d$ bit positions are $i_1, i_2, ..., i_d$.  
Then subset pairs $(A_{i_1}, B_{i_1}), (A_{i_2}, B_{i_2}),...,(A_{i_d}, B_{i_d})$ each differs by exactly $1$ (distinct) element
and hence can be reconciled using \autoref{alg:trivial_case}.

For this to happen, however, both Alice and Bob need to first know the values of $i_1, i_2, ..., i_d$, or the bit
positions where
$A[1..n]$ and $B[1..n]$ differ.
To this end, a naive solution is for Alice to send $A[1..n]$ to Bob and for Bob to compare 
it with $B[1..n]$.    
{\color{red}However, as mentioned earlier in \autoref{sec:smalld}, Alice can achieve the same goal by sending 
an ECC codeword $\xi_A$ that is much shorter than $A[1..n]$.  
The idea is that since $B[1..n]$ (which Bob already knows) 
can be viewed as a ``corrupted" 
(with $d$ bit errors in the positions $i_1, i_2, ..., i_d$) copy of 
$A[1..n]$, as long as the codeword $\xi_A$ is parameterized to correct at least $d$ bit errors, 
Bob can decode $B[1..n]\|\xi_A$ (the ``corrupted" message concatenated with the ECC codeword
of the ``uncorrupted" message)
to find out the $d$ bit error positions. } 

{\color{red}Although several ECC schemes are suitable for this purpose, we choose the BCH scheme for 
PBS because it results in near-optimal codeword length 
in the following sense: In the context of PBS, to ``correct up to $t$ bit errors", 
$\xi_A$ only needs to be $t \lceil\log (n + 1)\rceil$ bits long; 
even if Alice knew these $t$ bit positions precisely, specifying each bit position (to Bob) 
would require $\lceil\log n\rceil$ bits. 
BCH is also the choice of PinSkech~\cite{Dodis_PinSketch_2008} for the same reason.}

Once Bob decodes $B[1..n]\|\xi_A$ (\autoref{step:BCH_decode} in \autoref{alg:single_round_pbs}) to obtain $i_1,i_2,...,i_d$, 
Bob sends the \code{XOR} sums of the corresponding 
subsets $B_{i_1}$, $B_{i_2}$,$...$,$B_{i_d}$ to Alice (\autoref{alg1:sender} in \autoref{alg:trivial_case}).  Bob also needs to send the decoded ``bit error positions'' $i_1, i_2, ..., i_d$ to Alice (\autoref{step:send_xor} in \autoref{alg:single_round_pbs}), 
{\color{red} since Alice cannot obtain this information
by herself without knowing anything about $B[1..n]$.} 
In addition, for Alice to verify whether the set reconciliation is successfully completed (to be described next), Bob sends $c(B)$, a checksum of its set $B$, to Alice.  
\subsubsection{Verify the estimated set difference}\label{sec:checksum_verification}

Once Alice receives the ``bit error positions" and the corresponding \code{XOR} sums, 
she can recover the distinct elements each using \autoref{alg:trivial_case} to arrive at the estimated set difference
(\autoref{step:trivial_reconcile} in \autoref{alg:single_round_pbs}), which we denote as 
$\hat{D}$.   {\color{red}It is not hard to verify that in the ideal case this estimated set difference $\hat{D}$ is necessarily the same as the
actual set difference $A\triangle B$, so the unidirectional set reconciliation process is successfully completed.

However, the nonideal case can happen and when that happens $\hat{D}$ is in general not the same as $A\triangle B$.
Hence, Alice in general needs to verify whether $\hat{D}\stackrel{?}{=}A\triangle B$ after a round of set reconciliation process.}
Alice does so by checking an equivalent condition $A\triangle\hat{D}\stackrel{?}{=}B$ 
as follows.  She applies a
checksum function $c(\cdot)$ to $A\triangle\hat{D}$ and comparing (\autoref{step:terminate_condition} in \autoref{alg:single_round_pbs}) 
the resulting checksum $c(A\triangle\hat{D})$ with $c(B)$ that Alice received earlier from Bob. 
{\color{red}
We use as $c(\cdot)$ here the plain-vanilla summation function, with which the checksum of a set $S$ is the sum
of all elements (viewed as integers) modulo $|\mathcal{U}|$.   The length of such a checksum is $(\log |\mathcal{U}|)$ bits, the same as that of an element.
We use this checksum function for two reasons.  First, because it uses the `+' operation whereas the set reconciliation process (\autoref{alg:trivial_case}) involves a very different
operation (XOR), a 
false verification event is intuitively almost statistically uncorrelated\footnote{{\color{red}If we absolutely have to ensure any two such events to be provably strictly statistically uncorrelated, we can apply a one-way hash function to 
each element first and adding their hash values (viewed as integers) up instead, at a bit extra computation cost.}} with any reconciliation error (called an exception and to be described shortly) event, which makes the verification
step meaningful and effective to the maximum extent.  
Second, this checksum function can be incrementally computed. 
}

Using a 32-bit-long checksum (assuming $\log |\mathcal{U}| = 32$), the probability for Alice to mistakenly believe $A\triangle\hat{D}\!=\!B$ when the opposite 
(i.e., $\{A\triangle\hat{D}\!\neq\! B\}$) is true 
is only $O(10^{-12})$ for the following reason. 
The false verification event $\{A\triangle\hat{D}\! \neq\! B\}$ can happen only in the nonideal case, which happens with a probability of $O(10^{-2})$ (as we will show in \autoref{subsec:exception_hash_partition}).
Then, conditioned upon the event $\{A\triangle\hat{D}\! \neq\! B\}$ happening, 
the probability for their 32-bit-long checksums happen to be equal (i.e., $c(A\triangle\hat{D})\! =\! c(B)$) is only $2^{-32}\!\approx\! 2.3\!\times\! 10^{-10}$.
This $O(10^{-12})$ probability of incorrect verification should be acceptable in most applications.  

{\color{red}In applications in which correct verification absolutely has to be guaranteed ({\it e.g.,} bitcoin), 
additional built-in verification mechanisms, such as Merkle tree, are usually used, which can
reduce the probability of false verification to practically zero at no extra cost (to PBS).
For example, blockchain platforms Ethereum~\cite{Wood_Ethereumsecuredecentralised_2014} 
and Bitcoin~\cite{Nakamoto_Bitcoinpeerpeer_2019} both have Merkle tree~\cite{merkle87} based mechanisms for verifing the 
integrity and the consistency of transactions. 
A Merkle tree is a binary tree in which a parent node digitally certifies (verifies) its two children.  
In the cases of Ethereum and Bitcoin, each transaction corresponds to a leaf node of the Merkle tree that records 
the cryptographic hash value of the transaction, and 
each non-leaf node records the cryptographic hash value of its two children.  This way, the
root node digitally certifies the integrity and the consistency of all transactions.  
For mission-critical applications that do not have such an additional built-in verification mechanism, we can add one at a small cost. 
For example, we can compute and check $H(A\triangle\hat{D})\!\stackrel{?}{=}\!H(B)$, where $H$ is a 
one-way multiset hash function such as MSet-XOR-Hash~\cite{Clarke_IncrementalMultisetHash_2003}, 
at the additional cost of $O(\max\{|A|+d, |B|\})$ computation overhead and constant communication overhead. 
}

{\color{red}
In the case of PBS-for-small-$d$, the set reconciliation process will run as many rounds as it takes (to be explained in \autoref{sec:multi-rounds}) 
for the checksums of two sets being reconciled to eventually 
match each other;   
in the case of PBS-for-large-$d$, the same can be said about the set reconciliation process for each group pair (to be explained in~\autoref{sec:pbs-for-large-d}).
Hence, barring the false verification event, which as just explained happens with $O(10^{-12})$ probability when using only a 32-bit checksum or with practically zero probability when
using additional cryptographic verification techniques, the set reconciliation process (for both large and small $d$) 
guarantees to correctly reconcile $A$ and $B$ (and the respective referenced objects) when it halts. 
The formal proof can be found in~\refToAppendix{\autoref{app-sec:correctness}}{Appendix C in~\cite{Gong_pbs-tr_2020}}. 
With this understanding, for ease of presentation, we assume in the sequel that the checksum verification step will never produce a false verification.  

In~\Cref{subsec:exception_hash_partition,sec:BCH_exception}, we describe three types of exceptions may result in a $\hat{D}$ that is incorrect (not the same as $A\triangle B$).  
When that happens, the checksum verification step will not accept $\hat{D}$ as is, as just explained.  Hence, these exceptions will never result in 
an incorrect set reconciliation.  They can only delay the inevitable (eventual correct reconciliation of $A$ and $B$) 
by triggering additional rounds of set reconciliation process.
We note 
there is no need for PBS to determine which bin or bins cause the checksum verification step to fail in the current round,
because as we will show in~\autoref{sec:multi-rounds} such information is not used anywhere in the next round of set reconciliation operation.
}

\subsection{Exception Handling}\label{subsec:exception_hash_partition}

When the ideal case does not happen, some subset pairs may contain more than one distinct elements and cannot be successfully reconciled by \autoref{alg:trivial_case}. 
In this case, the checksum verification step will detect this event and trigger another round of PBS to reconcile the ``remaining'' distinct elements, as will be elaborated in \autoref{sec:multi-rounds}.  There are two types of exceptions that can possibly happen in such a subset pair, say $(A_i, B_i)$.

\smallskip
\noindent {\bf Type (\RN{1}) exception}:  $A_i\triangle B_i$ contains a nonzero even number of distinct elements.  In this case $A[i]=B[i]$
since the cardinalities of $A_i$ and $B_i$ are either both even or both odd.  The BCH codeword $\xi_A$ cannot detect this exception.
This exception happens with a small but nontrivial probability.  For example, when $d=5$ and $n=255$ (i.e., throwing $5$ balls each uniformly and randomly into $255$ bins), the probability that some bin has a nonzero even number of (in this case either $2$ or $4$) balls is roughly $0.04$. 


\smallskip
\noindent {\bf Type (\RN{2}) exception}: $A_i\triangle B_i$ contains an odd number (at least $3$) of distinct elements.  In this case, $A[i]\neq B[i]$. 
Bob will mistakenly believe that $(A_i, B_i)$ contains exactly one distinct element and try to recover it using \autoref{alg:trivial_case}. The ``recovered'' element $s$ is however the \code{XOR} sum of all distinct elements in $A_i\triangle B_i$ as explained at the end of \autoref{subsec:protocol0}.
We call this $s$ a fake distinct element.
This event happens with a tiny probability.  In the same example above ($d=5$ and $n=255$), 
the probability that some bin has an odd number of balls 
(in this case either $3$ or $5$) is only $1.52\times 10^{-4}$.  
This probability can be further reduced, thanks to the consistent nature of hash-partitioning, which provides us with a no-cost mechanism that can detect fake distinct elements 
{\color{red}(so that they will not be included in $\hat{D}$)} with high probability.

\begin{algorithm}
\caption{Check whether $s\in \mathcal{U}_i$}
\label{alg:verification}
\If{$h(s)\neq i$} {Discard the ``recovered'' element $s$;}
\end{algorithm}

As shown in \autoref{alg:verification}, the detection mechanism is simply to check whether $s\in \mathcal{U}_i$ (i.e., whether $h(s)\! =\! i$), 
a necessary condition 
for $s$ to be an element in $A_i\triangle B_i$.  
The conditional (upon a type (\RN{2}) exception happening) probability for a fake distinct element, which belongs to any of the $n$ sub-universes with equal probability $1/n$
since it is the \code{XOR} sum of multiple distinct elements in $A_i\triangle B_i$, to pass this check is only $1/n$.  In the same example above ($d\!=\!5$, $n\!=\!255$), 
this conditional probability
is roughly $3.9\!\times\! 10^{-3}$, and hence the probability for both a type (\RN{2}) exception to happen and the resulting fake distinct element to pass this check is 
$1.52 \times 10^{-4} \times 3.9 \times 10^{-3} \approx$ $6\times 10^{-7}$.
This mechanism only requires Alice to locally verify the ``recovered'' elements of \autoref{alg:trivial_case} and 
hence incurs no additional communication overhead. 

\subsection{Running PBS for Multiple Rounds}\label{sec:multi-rounds}


As mentioned earlier, when the ideal case does not happen, Alice and Bob cannot successfully reconcile their respective sets $A$ and $B$ 
{\color{red}in a single round}, 
and Alice can tell this situation from the checksum verification
step.  In this situation, Alice and Bob need to run additional rounds of \autoref{alg:single_round_pbs}, but with a different input set pair (than $(A, B)$) as follows. 
Let $\hat{D}_1$ be the estimated set difference Alice obtained in the first round.
In the second round, Alice and Bob try to reconcile their respective sets $A\triangle \hat{D}_1$ and $B$, from which Alice obtains another estimated difference (between $A\triangle \hat{D}_1$ and $B$) that we denote
as $\hat{D}_2$.  If the set reconciliation is still not successfully completed, Alice and Bob run a third round to try to reconcile sets $(A\triangle \hat{D}_1)\triangle \hat{D}_2$ and $B$.  This process continues
until the set reconciliation is successfully completed as verified by the checksum. 
{\color{red}The final output of the process, which is what Alice believes to be $A\triangle B$, is $\hat{D}_1\triangle\hat{D}_2\triangle\cdots\triangle \hat{D}_r$, 
where $r$ is the number of rounds this process runs 
and $\hat{D}_i$ for $i=1,2,\cdots,r$ is the estimated set difference in the $i^{th}$ round. 
}

In each subsequent round, a different and (mutually) independent hash function is used to perform the consistent hash partitioning of the two sets to be reconciled (e.g., $A\triangle \hat{D}_1$ and $B$ in the second
round), so that the same type (\RN{1}) and/or (\RN{2}) exceptions encountered in the previous round, which have so far prevented the set reconciliation from being successfully completed, can be avoided with 
overwhelming probability.  The use of independent hash functions in different rounds offers another significant benefit:  How the number of unreconciled distinct elements decreases one round after another (and 
eventually goes down to $0$) can now
be precisely modeled as a Markov chain, as will be elaborated in \autoref{sec:tail_bound}.

\subsection{BCH Encoding and Decoding}\label{sec:ecc-trick-bch}

{\color{red}
In this section, we describe the specific BCH encoding and decoding in PBS;  how this encoding differs from 
that for its usual application of communication over a noisy channel 
\refToAppendix{will be explained in~\autoref{app:bch}}{can be found in Appendix I in~\cite{Gong_pbs-tr_2020}}. 
Recall that in~\autoref{step:BCH_encode} of~\autoref{alg:single_round_pbs}, Alice sends, instead of the ``message'' $A[1..n]$ itself, its much shorter BCH codeword $\xi_A$ to Bob.  
We define the {\it error-correction capacity} of an ECC codeword as the maximum number of bit errors it can correct.
In the case of PBS-for-small-$d$, where $d$ is assumed to be known 
precisely beforehand,   
the error-correction capacity of $\xi_A$ is set to $d$
so the BCH decoding is always successful.  
However, as will be explained in~\autoref{subsec:parameterize-n-t}, when $d$ is large and the sets $A$ and $B$ each has to be partitioned into groups,
the number of ``bit errors" that occur to a group pair can exceed the error-correction capacity 
of the corresponding BCH codeword.  
In this case, a BCH decoding failure will happen and how to deal with its fallout will be explained in~\autoref{sec:BCH_exception}.

We now briefly explain what is involved for Bob to decode the BCH codeword $\xi_A$ against its local bitmap $B[1..n]$.
Here the only task is to figure out the ``bit error positions" (in which $A[1..n]$ and $B[1..n]$ differ).  To do so, Bob needs to 
invert a $d \times d$ matrix in which each matrix entry is an element of the finite field $GF(2^m)$ where $m \!=\! \lceil\log (n\! +\! 1)\rceil$.
In PBS, $n$ is always set to $2^m\!-\!1$ for some positive integer $m$ in BCH codes 
for achieving the maximum coding efficiency.  Hence, 
we drop ``floor" and ``ceiling" and consider $m \!=\! \log n$ in the sequel.
Normally such a matrix inversion would take $O(t^3)$ finite field operations over $GF(2^m)$.   However, since this matrix takes a special form called Toeplitz, 
it can be inverted in $O(d^2)$ operations over $GF(2^m)$ using the Levinson algorithm~\cite{levinson1946wiener}.  
}

\section{PBS for large \texorpdfstring{\larger\boldmath\MakeLowercase{$d$}}{d}}\label{subsec:protocol2}\label{sec:pbs-for-large-d}

In this section, we continue to assume that the number of distinct elements $d$ is precisely known in advance.
The PBS-for-small-$d$ scheme described in the previous section is no longer suitable when $d$ is very large, since its BCH decoding computational complexity is 
$O(d^2)$ finite field operations.  Instead, we first hash-partition sets $A$ and $B$ each into $g = d/\delta$ groups, and then apply PBS-for-small-$d$ to each of the $g$ group pairs.  
{\color{red}Here $\delta$ is the average number of distinct elements per group.  It is a tunable parameter, 
by which we can control the tradeoff between 
the communication and the computational overheads of PBS.  In general, the larger $\delta$ is, 
the lower the communication overhead and the higher the computational overhead are. 
We~\refToAppendix{will elaborate in~\autoref{app:vs-delta}}{have elaborated in Appendix J.2 in~\cite{Gong_pbs-tr_2020}} how 
$\delta$ controls this tradeoff.
Since $\delta$=5 appears to be a nice tradeoff point, we fix the value of $\delta$ at $5$ in this paper.}
Since each group pair contains on average $\delta = 5$ distinct elements, the BCH decoding computational complexity per group pair can be considered $O(1)$.  
As a result, the overall BCH decoding computational complexity is $O(d)$ for all $g = d/\delta$ group pairs.  We refer to this PBS-for-large-$d$ scheme as PBS in the sequel except in 
places where this abbreviation would result in ambiguity or confusion.

\subsection{How to Set Parameters {\texorpdfstring{\larger\boldmath\MakeLowercase{$t$}}{t}} and \texorpdfstring{\larger\boldmath\MakeLowercase{$n$}}{n}}\label{subsec:parameterize-n-t}

In PBS (i.e., PBS-for-large-$d$), we have to make some design decisions that we don't have to in PBS-for-small-$d$.  
One of them is how to set the error-correction capacities of the BCH codes used 
for each of the $g$ group pairs. 
Let $\delta_i$, $i=1,2,...,g$, be the number of distinct elements that group pair $i$ have.  
{\color{red}If we knew the precise values of $\delta_1$, $\delta_2$, 
..., $\delta_g$, we would simply set the BCH error-correction capacity 
for each group pair $i$, which we denote as $t_i$, to $\delta_i$.} This way, each BCH codeword is the shortest possible for the respective task, which minimizes the communication overhead of transmitting these codewords. 
In reality, we do not know the exact value of any $\delta_i$, since it is a random variable with distribution $Binomial(d, 1/g)$ thanks to the hash-partitioning (of $A$ and $B$ each into $g$ groups);  
we only know that $E[\delta_i] = d/g = \delta$  
but that does not help much.
In theory, we can measure $\delta_i$ using a (set difference) cardinality estimation protocol.  However, as will be shown in~\autoref{subsec:ToW}, to obtain such an estimate using the 
best protocol
would incur hundreds of bytes of communication overhead.  In comparison, the ``savings'' on the communication overhead that such an estimate would bring (for the corresponding group pair) 
is only tens of bytes, as we will elaborate next.

In PBS, we set an identical BCH error-correction capability $t$ for all $g$ group pairs.  It intuitively makes sense since random variables $\delta_1$, $\delta_2$, 
..., and $\delta_g$ are identically distributed.  Now the next question is ``How should we set this $t$?".   This is a tricky question because, 
on one hand, if $t$ is too large (say several times larger than $\delta$), then the total size of the BCH codewords is unnecessarily large, resulting in ``wastes" in communication overhead; 
but on the other hand, if $t$ is too small (say equal to $\delta$), then a large proportion of the BCH codewords cannot decode, resulting in considerable additional efforts and costs (i.e., 
``penalties'')
for reconciling the large proportion of affected group pairs.   In~\autoref{sec:tail_bound}, we propose an analytical framework that can be used to
identify the $t$ value that minimizes ``wastes + penalties'' (in~\autoref{sec:parameter_settings}).   This optimal $t$ value can range from $1.5\delta$ to $3.5\delta$  depending on how large this $d$ is.

Based on a similar rationale, we set another parameter for each group pair $i$ to the same value $n$:  Each group pair $i$ is to be partitioned into $n$ subsets, so that the parity bitmaps 
($A[1..n]$ and $B[1..n]$ in PBS-for-small-$d$) for all groups have the same length of $n$ bits.  This $n$ is also a tunable parameter (for optimal PBS performance), since 
the probability for the ideal case (of all distinct elements between a group pair being hashed to distinct subsets) to happen is a function of $n$ and $\delta = 5$.  
As will be elaborated in \autoref{sec:parameter_settings}, our 
analytical framework can also be used for the optimal tuning of $n$.

\smallskip
\noindent
{\bf Communication Overhead Per Group Pair.} 
Here we analyze the total communication overhead of the first round of PBS.  Since the vast majority of distinct elements are discovered and reconciled in the first round, 
as will be shown in \autoref{sec:piecewise_decodability}, it represents the vast majority (over $95\%$) of that over all rounds.  
For each group pair $i$, the communication overhead (of running PBS-for-small-$d$ on this pair) in the first round contains the following four components: (1) the BCH codeword that is $t\log n$ bits long; (2) the $\delta_i$ ``bit error locations'' whose total length is $\delta_i\log n$ bits; (3) the  $\delta_i$ \code{XOR} sums whose total length is $\delta_i\log |\mathcal{U}|$ bits; and (4) the checksum that is $\log |\mathcal{U}|$ bits long. Hence the average communication overhead of PBS per group pair in the first round is 

\begin{equation}\label{eq:communication_overhead}
t\log n + \delta\log n +\delta \log |\mathcal{U}|+\log |\mathcal{U}|
\end{equation}

\subsection{Exception Handling on BCH Decoding}\label{sec:BCH_exception}

Recall that in PBS-for-small-$d$ we need to handle two types of exceptions:  type (\RN{1}) and type (\RN{2}). 
In PBS-for-large-$d$, we have another exception to worry about.
{\color{red}This exception arises when the number of bit positions where bitmaps $A$ and $B$ in~\autoref{alg:single_round_pbs} 
differ is larger than $t$, the universal BCH error-correction capability (for every group pair).
When this exception happens,} the BCH decoding would fail (when executing \autoref{step:BCH_decode} in \autoref{alg:single_round_pbs}) 
and the decoder would report a failure.
With $t$ appropriately parameterized as explained earlier, this exception should happen with a small probability to any group pair. 
For example, when $d=\num{1000}$, $\delta=5$ (so that $g=d/\delta=200$), and $t$ is set to the 
optimal value of $13$ ($=2.6\delta$), the probability for this exception to happen to any group pair
is only $6.7\times 10^{-4}$. 

To handle this type of exceptions, we further hash-partition each trouble-causing group pair (whose BCH decoding has failed) 
into $3$ sub-group-pairs and reconcile each of them using PBS-for-small-$d$. 
With this three-way split, with an overwhelming probability, each sub-group-pair should contain no more than $t$ distinct elements and its BCH decoding operation should be 
successful in the next round.  For example, when $\delta=5$ and $t=13$ (same as in the example above), 
the conditional (upon the event $\{\delta_i>t\}$, which happens with probability $6.7\times 10^{-4}$ as 
just explained) probability for any sub-group-pair to contain more than $t$ distinct elements is only $9.5\times 10^{-10}$. 
{\color{red} We use a three-way split here because a two-way split would result in a much higher conditional probability for this event:
In the same example above ($\delta=5$ and $t=13$), this conditional probability becomes $0.0012$.} 
All said, if necessary, a trouble-causing sub-group-pair will be further split three-way.

{\color{red}
As explained earlier, the ultimate gatekeeper for ensuring the correctness of set reconciliation is the checksum verification step, which
in this case (of large $d$) is applied to each group pair.  BCH decoding exceptions alone, or in combination with type (\RN{1}) or (\RN{2}) exceptions, may only 
delay the inevitable eventual correct reconciliation of $A$ and $B$, as long as a  false checksum verification event does not happen.
}

\subsection{Multi-round Operations}\label{sec:multiround_large_d}

In PBS, the set reconciliation processes of the $g$ group pairs are independent of each other.  
Each group pair runs as many rounds of PBS-for-small-$d$ as needed to reconcile all distinct elements between them. 
{\color{red}Almost every set reconciliation scheme is designed and parameterized to provide the performance guarantee that 
the reconciliation process is successfully completed, in the sense all distinct elements are correctly reconciled, with at least a target probability $p_0$.}   
In PBS, this guarantee will involve an additional 
parameter $r$ that is the target number of rounds the scheme is allowed to run to reach this target success probability.     More precisely, 
the multi-group-pair multi-round operation of PBS must, with a probability that is at least $p_0$, be successfully completed 
in $r$ rounds.  

Let $R$ be the number of rounds it takes for all $g$ group pairs, and hence the set pair, to be successfully reconciled.
This guarantee can then be succinctly written as $Pr[R\leq r] \ge p_0$.  Intuitively, we can always provide this guarantee by making the values of the two key parameters
$n$ (the size of the parity bitmap) and $t$ (the BCH error-correction capacity) very large, but doing so would 
result in a high communication overhead.  This apparent tradeoff leads us to study the following parameter optimization problem:
Among all parameter settings of $n$ and $t$ that can 
guarantee $Pr[R\leq r] \ge p_0$, which one results in the smallest communication overhead?

To tackle this optimization problem, we need to first analyze the multi-group-pair
success probability $Pr[R\leq r]$.  
The latter boils down roughly (but not exactly as 
\refToAppendix{we will explain in \autoref{sec:times2bound}}{we have explained in Appendix F in~\cite{Gong_pbs-tr_2020}}) to analyzing the following single-group-pair success probability.
Consider a single group pair that have
$x$ distinct elements between them before the first round starts.    For the moment, we assume $x \le t$
so that we do not have to worry about the BCH decoding failure.   Our problem is to derive the formula for the probability of the 
following event that we denote as $\{x\!\xrsquigarrow{r}\! 0\}$:  All the $x$ 
distinct elements, and hence the pair, are successfully reconciled in no more than $r$ rounds.   
Solving this problem is the sole topic of \autoref{sec:tail_bound}.  

\section{Analytical Framework}\label{sec:tail_bound}\label{sec:single_markov}

In this section, we derive a Markov-chain model for analyzing the aforementioned single-group-pair success probability $Pr[x\!\xrsquigarrow{r}\! 0]$.  
We will show next that, 
under this model, the initial state of the Markov chain is state $x$ (distinct elements), 
and each (set reconciliation) round triggers a state transition.  
Hence, the event $\{x\!\xrsquigarrow{r}\! 0\}$ corresponds to the Markov chain reaching the ``good" 
state $0$ (distinct elements left) within $r$ transitions.  
Suppose the transition probability matrix of this Markov chain is $M$. 
The formula for computing the probability of this event is simply

\begin{equation}\label{eq:single_success_prob}
Pr[x\!\xrsquigarrow{r}\! 0] = \big{(}M^r\big{)}(x, 0)
\end{equation}

Here $\big{(}M^r\big{)}(x, 0)$ is the element at the intersection of the $x^{th}$ row and the 
$0^{th}$ column in the matrix $M^r$ ($M$ to the power $r$).  Note that, 
without this Markov-chain model, $Pr[x\!\xrsquigarrow{r}\! 0]$ is hard to compute except in some
special cases such as when $r = 1$ (the probability, derived earlier in \autoref{sec:hash_partition}, for the ideal case of $x$ balls thrown ending up in $x$ distinct bins to happen).

{\color{red}We now describe this Markov-chain model.}  
How the number, starting at $x$ before the first round, of yet unreconciled distinct elements between the group pair 
decreases one round after another, and 
eventually goes down to $0$, can 
be precisely modeled as a Markov chain as follows.   
As described in \autoref{sec:multi-rounds}, in the first round, each of the $x$ balls (distinct elements) is thrown uniformly and randomly (by the hash function $h$) 
into one of the $n$ bins (subset pairs).  If a ball ends up in a bin that contains no other balls, the corresponding distinct element can be 
successfully reconciled using \autoref{alg:trivial_case}.  We call this ball a ``good'' ball, since it does not have to be thrown again in later rounds, and for the 
modeling purpose call
this bin a ``good'' bin (just for this round).  
If a ball ends up in a bin that has other balls, which corresponds to a type (\RN{1}) or type (\RN{2}) exception discussed earlier in \autoref{subsec:exception_hash_partition},
the corresponding distinct element cannot be reconciled in this round.   We call this ball a ``bad'' ball, since it has to 
be thrown again in the second round (in the hope of making it ``good'' this time), and for the modeling purpose call this bin a ``bad'' bin (just for this round).  

As described in \autoref{sec:multi-rounds}, the ``bad'' balls (if any) that remain after the first round will be thrown again in the second round, the ``bad'' balls (if any) 
that remain in the second round will be thrown again in the third round, and so on.  
Let $D_k$, $k=1,2,\cdots$, be the number of balls 
that remain ``bad'' (yet unreconciled distinct elements) after the $k^{th}$ round.  
Let $D_0 = x$ be the number of balls to be thrown at the beginning (i.e., right before the first round).   
Then $\{D_k\}_{k=0}^\infty$ is a Markov chain
for the following reason.   Since a different and mutually independent hash function is used in each round, 
the random variable $D_{k}$, 
which is the number of balls that remain ``bad'' after the $k^{th}$ round, depends only on $D_{k-1}$, the number of balls thrown in the $k^{th}$ round, and is conditionally (upon $D_{k-1}$) independent of the history $D_0, D_1, D_2, \cdots, D_{k-2}$. 

The states of this Markov chain are $i = 0, 1, 2, \cdots$.  Each state $i$ means that there are $i$ ``bad'' balls to be thrown at the beginning of a round.  
The (desired) termination state of this Markov chain is state $0$ (``bad'' balls remaining).
For both notational convenience and confusion minimization, we shift each row/column index down by 1.  In other words, we refer to the 
first row/column as the $0^{th}$ row/column, the second as the first, and so on.   This way, the matrix element $M(i, j)$ (at the intersection of the $i^{th}$ 
row and the $j^{th}$ column under the new index numbering convention) corresponds to the probability for 
the Markov chain to move from state $i$ (``bad'' balls thrown) to state $j$ (``bad'' balls remaining).  

{\color{red}In the interest of space, we leave out here our discussions on the preciseness of this Markov-chain model and on how the matrix $M$ is computed.
They can be found in~\refToAppendix{\autoref{sec:preciseness} and~\autoref{sec:compute_M}}{Appendices D and E in~\cite{Gong_pbs-tr_2020}}.}

\section{Applying the framework}\label{sec:application_analysis}
{\color{red}Knowing the Markov-chain model and how to compute its transition matrix $M$}, we are now 
ready to tackle the aforementioned parameter optimization problem in \autoref{sec:parameter_settings} and 
study two other related parameterization and design questions in \autoref{sec:vary_r} and \autoref{sec:piecewise_decodability} respectively.

\subsection{Parameter Optimization}\label{sec:parameter_settings}

Recall that our optimization problem is to find the optimal parameter settings of $n$ and $t$ that 
guarantee $Pr[R\leq r] \ge p_0$ yet result in the smallest communication overhead.  
Recall that our original goal is to analyze the overall (for all $g$ group pairs) success probability 
$Pr[R\leq r]$.
{\color{red}\refToAppendix{In \autoref{sec:times2bound}, we will show}{In Appendix F in~\cite{Gong_pbs-tr_2020}, we have shown} 
that $Pr[R\leq r]$ is hard to calculate exactly, but can be tightly lower-bounded by 
$1\!-\!2(1\!-\!\alpha^g)$, where 
$\alpha\triangleq\sum_{x=0}^t Pr[X\!=\!x]\cdot Pr[x\!\xrsquigarrow{r}\! 0]$ is a slightly underestimated success probability for any group pair, 
$g$ is the number of group pairs, and $t$ is the error-correction capacity.
Here the random variable $X$ is distributed as $Binomial(d, 1/g)$. 
}

\smallskip
\noindent
{\bf Minimize Communication Overhead.}
Armed with this rigorous lower bound on the overall success probability $Pr[R \le r]$, we can now formulate our optimization problem of parameterizing PBS to guarantee $Pr[R \le r] \ge p_0$ while minimizing the average 
communication overhead as follows.


\begin{align}
\mbox{minimize} \quad
&
t\log n+\delta\log n \nonumber
\\
\mbox{subject to} \quad 
&
1-2(1-\alpha^g(n,t))\geq p_0,\quad n,t\in \mathbb{N}^+ \nonumber
\end{align}

The objective function $t\log n+\delta\log n$ (as a function of $n$ and $t$) here is the non-constant part of the average communication overhead per group pair in the first round as shown in Formula \autoref{eq:communication_overhead}.
It is an appropriate objective function because it is exactly $1/g$ of the average communication overhead for all $g$ group pairs in the first round, and as explained earlier and will be confirmed later, the first round incurs over
$95\%$ of the total communication overhead. 
In the constraint, we replace $Pr[R\leq r]$ by its lower bound $1\!-\!2(1\!-\!\alpha^g)$ and write $\alpha$ 
as $\alpha(n, t)$ to emphasize it is a function of $n$ and $t$, when $r$ is considered a constant. 
In this optimization problem, $g$ (in the constraint) is a constant, since Alice and Bob both know $d$ (by our assumption thus far), and $g = d/\delta$.
Here $\delta$ is the average number of distinct elements per group, which we set to $5$ in PBS.
Hence there are only two variables involved in this optimization problem: $n$ and $t$.

This optimization problem is not as daunting as it might appear, since there are only a few meaningful value 
combinations of $n$ and $t$ for two reasons. 
{\color{red}
As mentioned earlier in~\autoref{sec:ecc-trick-bch}, $n$ is always set to $2^m\!-\!1$ for some integer $m$ in PBS.}
Also, $n$ cannot be too small, since otherwise the ideal case (of $x$ ``balls" landing in $x$ distinct ``bins") cannot happen with high probability. 
The possible $n$ values are hence narrowed down to $\{63, 127, 255, 511, 1023, 2047\}$ in practice.
Second, the BCH error-correction capacity $t$ needs to be set to between $1.5\delta$ and $3.5\delta$, as explained in \autoref{subsec:protocol2}, to strike a nice tradeoff between the probability of BCH decoding failure and the increase in BCH codeword length.

Our optimization procedure is simply to compute, for each of the 100 or so value combinations of $n$ and $t$, the corresponding values of the lower bound $1-2(1-\alpha^g(n,t))$ 
(of $Pr[R\leq r]$) and the objective function $t\log n+\delta\log n$.  
Then among all such value combinations that can guarantee $Pr[R\leq r] \ge p_0$, we pick the one that results in the smallest objective function value. {\color{red}We provide a detailed example in~\refToAppendix{\autoref{app:exam-param-op}}{Appendix H in~\cite{Gong_pbs-tr_2020}} to illustrate how this procedure works with the following parameter settings:
$d\!=\!\num{1000}$ distinct elements, $\delta\!=\!5$ (so that $g\!=\!200$ groups), $r\!=\!3$ rounds, and
the target success probability $p_0\!=\!99\%$.
}

\subsection{What If The Target {\texorpdfstring{\larger\boldmath\MakeLowercase{$r$}}{r}} Changes?}\label{sec:vary_r}



Intuitively, when the target number of rounds $r$ becomes smaller, it becomes more costly, in terms of both the communication and the computational (for BCH decoding) overheads, to provide 
the success probability 
guarantee $Pr[R\!\leq\! r]\!\geq\! p_0$.  Intuitively, this is because $n$ and $t$ have to be larger so that in each group pair the ideal case happens and the BCH decoding succeeds with 
higher probabilities respectively. 
{\color{red}In this section, we perform a quantitative study of this tradeoff, using 
the same example above with $p_0=99\%, d=\num{1000}$ as that used in~\autoref{sec:parameter_settings}. }
For each $r \in \{1, 2, 3, 4\}$, we compute the optimal $(n,t)$ value combination using the optimization procedure described above, and the corresponding
optimal (minimum) average communication overhead {\it per group pair}.  

{\color{red}The optimal communication overheads per group pair are \numlist[list-pair-separator = {, and }]{591;402;318;288} 
bits when $r=$\numlist[list-pair-separator = {, and }]{1;2;3;4} respectively,
which confirms our earlier intuition that the larger the $r$ is, 
the smaller the optimal communication overhead is.} 
It also shows that $r = 3$ is a sweet spot:  The communication overhead per group 
pair drops sharply from when $r = 1$ (591 bits) to 
when $r = 2$ (402 bits) and from when $r = 2$ to when $r = 3$ (318 bits), but drops only 
slightly from when $r = 3$ to when $r = 4$ (288 bits).  
We have found that $r = 3$ is in general a sweet spot whenever the target success 
probability $p_0$ is relatively high, such as $p_0=99\%$ and $p_0=99.58\%$ $(239/240)$ that will be used in our evaluation.
Hence we set $r$ to 3 in this paper.  
For smaller $p_0$ values, however, $r = 2$ or 
even $r = 1$ can become a sweet spot, as long as $d$ is not gigantic (say tens of millions and beyond). 

\subsection{Analysis on ``Piecewise Reconciliability''}\label{sec:piecewise_decodability}



In this section, we perform a quantitative study of what portion of the distinct elements are expected to be reconciled by PBS in the first round, 
in the second round, and so on, again using our Markov-chain model.
The study confirms our earlier claim that the 
vast majority ($>95\%$) of the distinct elements are reconciled, and hence most of the communication overhead is
incurred, in the first round.

{\color{red}The detailed theoretical analysis for this quantitative study is omitted here in the interest of 
space, which can be found in~\refToAppendix{\autoref{app-sec:analysis-piecewise}}{Appendix G in~\cite{Gong_pbs-tr_2020}}. 
Using this analysis, we obtain that the expected proportions of the $d$ distinct elements 
that are reconciled in the first, second, third, and fourth round are 
$0.962,0.0380,3.61\times 10^{-4}$, and $2.86\times 10^{-6}$ respectively 
under the optimal parameter settings ($n\!=\!127$, $t\!=\!13$) for the instance used twice above 
(with $d\!=\!\num{1000}$, $r\!=\!3$, $\delta\!=\!5$ and $p_0\!=\!0.99$). 
That confirms our earlier claim.}
Through experiments, we have found that this claim holds in general under a wide range of $d$ and $p_0$ values.  

\section{Estimate \texorpdfstring{\larger\boldmath\MakeLowercase{$d$}}{d}}\label{subsec:ToW}

We have so far assumed that $d$ is precisely known. In reality, $d$ is not known a priori in most applications. 
In this case, 
Alice and Bob need to first obtain a relatively accurate estimate of $d$. 
To this end, 
we propose a new set difference cardinality estimator that is based on 
the celebrated 
Tug-of-War (ToW) sketch~\cite{AlonMS99}.

\subsection{The ToW Estimator}\label{subsec:tow-estimator}

The ToW sketch was originally proposed in~\cite{AlonMS99} for a subtly related but very different application: 
to estimate $F_2$, the second frequency moment of a data stream. 
We discover that ToW can also be used for estimating the 
set difference cardinality $d$ as follows.  Given a universe $\mathcal{U}$, let $\mathcal{F}$ be a family of four-wise independent hash functions, each of which maps 
elements in $\mathcal{U}$ to $+1$ or $-1$ each with probability 0.5. The ToW sketch of a set $S\subset \mathcal{U}$,  
generated using a hash function 
$f\in\mathcal{F}$, is defined as $Y_f(S)\triangleq \sum_{s \in S}f(s)$, 
the sum of the hash values of all elements in $S$. 
Using the same analysis derived in~\cite{AlonMS99}, 
we can prove that $\hat{d} = \big{(}Y_{f}(A)\!-\!Y_{f}(B)\big{)}^2$
is an unbiased estimator for $d=|A\triangle B|$,  
as long as $f$ is drawn uniformly at random from $\mathcal{F}$. 
The variance of this estimate is $(2d^2\!-\!2d)$.  
{\color{red}The proof for the unbiasedness of this estimator and the calculation for its variance can 
be found in~\refToAppendix{\autoref{app-sec:mean-var-proof}}{Appendix A in~\cite{Gong_pbs-tr_2020}}.} 
For notational convenience, we drop the subscript $f$ from $Y_f$ and add a different subscript to $Y$ in the sequel.

The estimate obtained from a single sketch is usually not very accurate. 
To achieve high estimation accuracy, multiple sketches, generated using independent hash functions, can be used. 
Suppose $\ell$ such sketches, which we name $Y_1$, $Y_2$, ..., $Y_\ell$, are used.   
The ToW estimator using these $\ell$ sketches 
is given by  
$\hat{d}\!=\!\big{(}\sum_{i=1}^\ell (Y_i(A)-Y_i(B))^2\big{)}/\ell$.
The variance of $\hat{d}$ is 
$(2d^2\!-\!2d)/\ell$, which is $\ell$ times smaller than if only a single ToW sketch is used.

\def\estspace{336}
\smallskip
\noindent
{\bf Space Complexity.} 
Each ToW sketch for any set $S$ is an integer within the range $[-|S|,|S|]$, and is hence  
at most $\log(2|S|+1)$ bits long.  Therefore, the space complexity of the ToW estimator using $\ell$ sketches is $\ell\cdot\log(2|S|+1)$ bits. 
{\color{red}We use $\ell$=128 totaling \SI{\estspace}{bytes} in PBS to achieve an appropriate level of estimation accuracy that we will elaborate next.}



\subsection{Use The ToW Estimator in PBS}\label{subsec:combine-with-pbs}

The ToW estimator is to be used by PBS, or by any other set reconciliation algorithm that needs this step, at the very beginning (before the 
reconciliation process starts), as follows. 
Alice sends the $\ell\!=\! 128$ ToW sketches of set $A$ to Bob.  Upon receiving these $\ell$ ToW sketches, Bob computes  
$\hat{d}$ as shown above and sends $\hat{d}$ to Alice.   
Both Alice and Bob then conservatively assume that the actual $d$ is $1.38\hat{d}$ and compute the optimal
$n$ and $t$ values (described in~\autoref{sec:parameter_settings}) accordingly.  
We use $\gamma = 1.38$ here, because it is found (through Monte-Carlo simulations) to be the smallest $\gamma$ value to guarantee that 
$Pr[d \le \gamma\hat{d}]\ge 99\%$ for the ToW estimator using 128 sketches.  
Using more (than 128) sketches allows
$\gamma$ to be smaller, but (128, 1.38) appears to strike a nice tradeoff according to our simulations. 


In our evaluation to be described in~\autoref{sec:evaluation}, we assume that $d$ is not known a priori. 
Like in PBS, we use the ToW estimator 
with 128 sketches with a total cost of \SI{\estspace}{bytes} 
also for two of our ``competitors" PinSketch and D.Digest, because, 
{\color{red}\refToAppendix{as to be explained next in~\autoref{subsec:other-estimators}}{as have been explained in Appendix B in~\cite{Gong_pbs-tr_2020}}}, the ToW estimator is the most 
space-efficient among all existing estimators. 
In calculating the communication overheads of all three of them (PBS, PinSketch, D.Digest),
this overhead of \SI{\estspace}{bytes} 
is excluded. 
For a fair comparison, we subtract this   
amount (\SI{\estspace}{bytes}) from the communication overhead of another competitor Graphene, 
as Graphene does not require an estimator.

\section{Related Work}\label{sec:related}

In this section, we provide a brief survey of existing set reconciliation algorithms. 
As mentioned in~\autoref{sec:intro_prob_formulation}, in describing and comparing them with PBS, 
we only consider the unidirectional set reconciliation in which 
Alice learns $A\triangle B$. 

Bloom filters (BF)~\cite{Bloom_Space/TimeTradeoffs_1970} can be used to construct a crude set reconciliation scheme as follows.  
First, Alice and Bob exchange BFs for sets $A$ and $B$.  Upon receiving the BF (for $B$) from Bob, 
Alice can obtain an estimate, denoted as $\widehat{A\!\setminus\! B}$, of the set $A\!\setminus\! B$ 
by checking each element in $A$ against this BF. 
Note that $\widehat{A\!\setminus\! B}$ is in general an underestimate:
$\widehat{A\!\setminus\! B}$ may not contain all elements in $A\!\setminus\!B$, because this BF may produce false positives  
that each suggests an element is in $B$ when it is not. 
Similarly, Bob can obtain $\widehat{B\!\setminus\! A}$ and sends it to Alice, from which Alice can infer an underestimate (of $A\triangle B$)
$\widehat{A\triangle B}\!=\!\widehat{A\!\setminus\! B}\bigcup\widehat{B\!\setminus\! A}$. 
Set reconciliation solutions that build on and extend this BF-based technique, including~\cite{Byers_Informedcontentdelivery_2004,Guo_SetReconciliationvia_2013,Luo_SetReconciliationCuckoo_2019},
all suffer from this underestimation problem, and hence 
{\color{red}are only suitable} 
for few applications that do not require perfect data
synchronization.

As mentioned earlier, most exact set reconciliation algorithms
are based on either invertible Bloom filters (IBF)~\cite{Goodrich_InvertibleBloomLookup_2011} or error-correction codes (ECC). 

\smallskip
\noindent
{\bf IBF-Based Algorithms.}
In an IBF, each element (from a set) is inserted into $k$ cells indexed by $k$ independent hash functions. Whereas each cell is a single 
bit in a BF, it has three fields in an IBF, each of which requires a single word of length $\log |\mathcal{U}|$. Therefore, IBFs are much more powerful than BFs: The set difference $A\triangle B$ of sets $A$ and $B$ can be recovered from the ``difference'' of their IBFs using a ``peeling process'' 
similar to that used in the decoding algorithms for some erasure-correcting codes, such as Tornado codes~\cite{luby1998tornado}.  
For this decoding process to succeed with a high enough probability, 
IBF-based solutions, such as 
Difference Digest (D.Digest)~\cite{Eppstein_WhatsDifference?Efficient_2011}, have to use roughly $2d$ cells.
This translates into a communication overhead of roughly $6d\log|\mathcal{U}|$, or
6 times the theoretical minimum. 


A recent solution called Graphene~\cite{Ozisik2019} reduces the high communication overhead of IBF-based solutions by augmenting it with  BFs.
Here, we only describe its simplest version (Protocol I in~\cite{Ozisik2019}) 
that works only for the special case of $B\!\subset\! A$. 
Its basic idea is for Alice to first obtain $\widehat{A\!\setminus\! B}$, an underestimate of of $A\!\setminus\! B$, by querying the BF for the set $B$ as described above, and then
recover only the ``missing" part $(A\!\setminus\! B) \!\setminus\! (\widehat{A\!\setminus\! B})$ using an IBF.   When the BF is configured to have a reasonably low false 
positive rate say $\epsilon$, 
the IBF needs only to ``encode" the roughly $\epsilon d$ ``missing" distinct elements rather than all the $d$ elements in $A\!\setminus\! B$, resulting in savings of 
``$O((1-\epsilon)d)$" in the size of IBF.    
In general, for this $\epsilon$ to be reasonably low (say meaningfully away from $1$), 
the size of the BF has to be $O(|B|)$ with a nontrivial constant factor~\cite{Bloom_Space/TimeTradeoffs_1970}. 
However, when $|B| \!\gg\! d$, which as explained earlier is often the case in most applications, 
the savings of ``$O((1\!-\!\epsilon)d)$" in the IBF size is no longer worth the $O(|B|)$ cost of the BF;
in this case Graphene drops the BF and degenerates to an IBF-only solution.    
For this reason, Graphene is more communication-efficient than other IBF-based solutions only when
$d$ is sufficiently large with respect to $|B|$.  Furthermore, this efficiency grows with $d$, as we will show 
in~\autoref{subsec:vs-graphene}.

{\color{red}
\smallskip
\noindent
{\bf ECC-Based Algorithms.}
The basic ideas of 
ECC-based algorithms~\cite{Minsky2003,Dodis_PinSketch_2008,Naumenko_ErlayEfficientTransaction_2019,Karpovsky_eccreconciliation_2003} 
are similar to that of PinSketch~\cite{Dodis_PinSketch_2008}. 
Given a universe $\mathcal{U}$ 
in which each element is assigned an index between $1$ and $|\mathcal{U}|$, 
PinSketch 
encodes each set $S\!\subset \!\mathcal{U}$ as a 
$|\mathcal{U}|$-bit-long bitmap $S[1..|\mathcal{U}|]$: $S[i]$ (the $i^{th}$ bit of $S[1..|\mathcal{U}|]$) 
is equal to $1$ 
if the element, whose index is $i$, is contained in $S$; otherwise, $S[i]=0$. 
For example, when $|\mathcal{U}| \!=\! 2^{32}$ as assumed in most existing works, 
the bitmap encoding for any set in PinSketch is $2^{32}$ bits long.
In contrast, in PBS 
the size $n$ of a bitmap depends only on $d$ (in PBS-for-small-$d$) or $\delta$ (in PBS-for-large-$d$),
and not on the size of the universe or the cardinality of the group the bitmap encodes, and is hence much shorter.  
For example, as shown in the example in~\autoref{subsec:exception_hash_partition}, $n \!=\! 255$ is large enough for $d \!=\! 5$.  

In PinSketch the $d$ distinct elements in $A\triangle B$ are ``indexed" by the $d$ bit positions in which the 
two $|\mathcal{U}|$-bit-long bitmaps encoding $A$ and $B$ respectively differ.  Like in PBS, these $d$ bit locations
can be learned by letting Alice send BoB a BCH codeword encoding $A$'s bitmap.   However, whereas the length of
BCH codeword in PBS is $d\log n$ or $\log n$ per distinct element, 
that in PinSketch is $d\log |\mathcal{U}|$, or $\log |\mathcal{U}|$ per distinct element.
Hence, the BCH codeword is typically 3 to 4 times longer ($\log |\mathcal{U}|$ = 32 bits in the example above) 
in PinSketch than in PBS ($\log n \!=\! 8$ bits in the example above), a fact we will use in~\autoref{subsec:vs-pinsketchwp}.


As mentioned earlier, ECC-based algorithms suffer from a much higher decoding 
computational complexity of at least $O(d^2)$. 
In~\cite{Minsky_Practicalsetreconciliation_2002}, a partition-based solution was proposed 
to reduce this computational complexity to $O(d)$, but in a different manner than the partitioning in PBS. 
This solution contains an ECC-based algorithm,  
called {\it BASIC-RECON}, that can 
reconcile a small number of distinct elements, just like what PBS-for-small-$d$ does in PBS.        
This solution   
recursively two-way partitions sets $A$ and $B$ each until each partition pair can 
be successfully reconciled by {\it BASIC-RECON}. 
Hence this solution requires $O(\log d)$ rounds of message 
exchanges, which is generally much larger than that in PBS. 
}

\section{Performance Evaluation}\label{sec:evaluation}
In this section, we evaluate the performance of PBS, and compare 
it against three state-of-the-art algorithms that we have 
described in detail in~\autoref{sec:related}: PinSketch~\cite{Dodis_PinSketch_2008}, Difference Digest (D.Digest)~\cite{Eppstein_WhatsDifference?Efficient_2011}, 
and Graphene~\cite{Ozisik2019}. 
{\color{red}In~\autoref{subsec:vs-pinsketchwp}, we apply the partitioning technique used in PBS to PinSketch to 
reduce its decoding computational complexity and compare PBS against it.}
Our evaluation is mainly focused on 
two performance metrics: {\it communication overhead} and {\it computational overhead}. 
The former is measured by the total amount of data 
transmitted between Alice and Bob to allow Alice to 
learn $A\triangle B$. 
The latter includes both encoding and decoding times. 


%
%
%
%
The evaluation shows conclusively that 
PBS strikes a much better tradeoff between communication and computational overheads than all three algorithms. 
It has a communication overhead much lower than IBF-based techniques such as D.Digest and Graphene, 
and only slightly higher than PinSketch, whose computational overhead is much larger. 
In addition, PBS has the lowest computational overhead among all four algorithms. 
\smallskip
\noindent
{\bf Experiment Setup.}
Our evaluation uses a key space (universe) $\mathcal{U}$ of all $32$-bit binary strings.  
In other words, the (hash) signature length is $\log|\mathcal{U}|\!=\!32$. 
%
%
Like in~\cite{Eppstein_WhatsDifference?Efficient_2011}, 
all set pairs
are created as follows. 
First, elements in $A$ are drawn from 
$\mathcal{U}$ 
uniformly at random without replacement. 
A certain number (more precisely, $|A|\!-\!d$) 
of elements in $A$ are then sampled also 
uniformly at random without replacement  
to make up set $B$ so that the set difference $A\triangle B$ 
contains exactly $d$ elements. 

{\color{red}In all experiments, we fix the cardinality of $A$ at $10^6$ and 
let the value of $d$ vary from $10$ to $10^5$.}
%
For each value of $d$, we create a set of \num{1000} mutually independent 
instances of $(A,B)$.  
Each point in each plot  
is the average of 
\num{1000} experimental results on such a set of \num{1000} instances. 
All experiments were performed on a workstation with an Intel Core i7-9800X processor 
running Ubuntu 18.0.4. 
\smallskip
\noindent
{\bf Implementations.}
We implement PBS in C++. We use the \texttt{xxHash} library~\cite{Collet_xxHashExtremelyfast_} for generating all hash functions in PBS, including those in the ToW estimator. 
The Minisketch library~\cite{Wuille_Minisketchoptimizedlibrary_2019}, released by the authors of~\cite{Naumenko_ErlayEfficientTransaction_2019}, is used for the BCH encoding and decoding in both PBS and PinSketch~\cite{Dodis_PinSketch_2008}. 
As the authors of D.Digest~\cite{Eppstein_WhatsDifference?Efficient_2011} have not 
released their source code, we implement it using the open-source code of IBFs in C++ released by 
the authors of~\cite{Ozisik2019}. 
%
{\color{red}For evaluating Graphene~\cite{Ozisik2019} fairly, we have made the following revision 
to the source code provided by the authors of~\cite{Ozisik2019} to make it as computationally efficient as possible.  
The original source code was written in Python, with the most computationally expensive part implemented in C++ with a 
Python wrapper.  We have rewritten all Python code in it using C++.}

\begin{figure*}[t] 
\centering
\begin{minipage}{\textwidth}
\begin{subfigure}[t]{0.25\textwidth}
\centering
\includegraphics[width=0.99\textwidth]{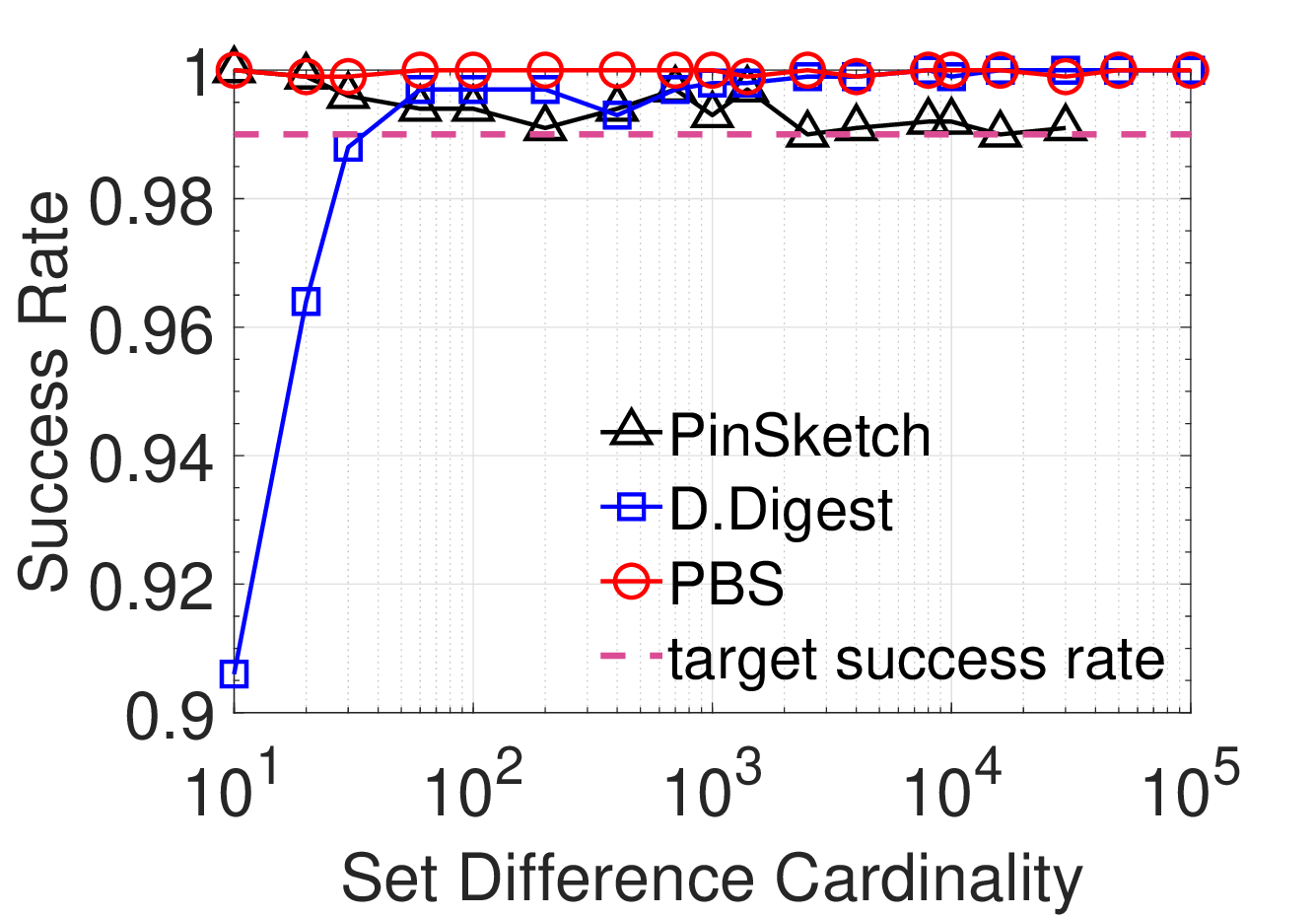}
\caption{Success rate}
\label{fig:success-rate-99}
\end{subfigure}%
\begin{subfigure}[t]{0.25\textwidth}
\centering
\includegraphics[width=0.99\textwidth]{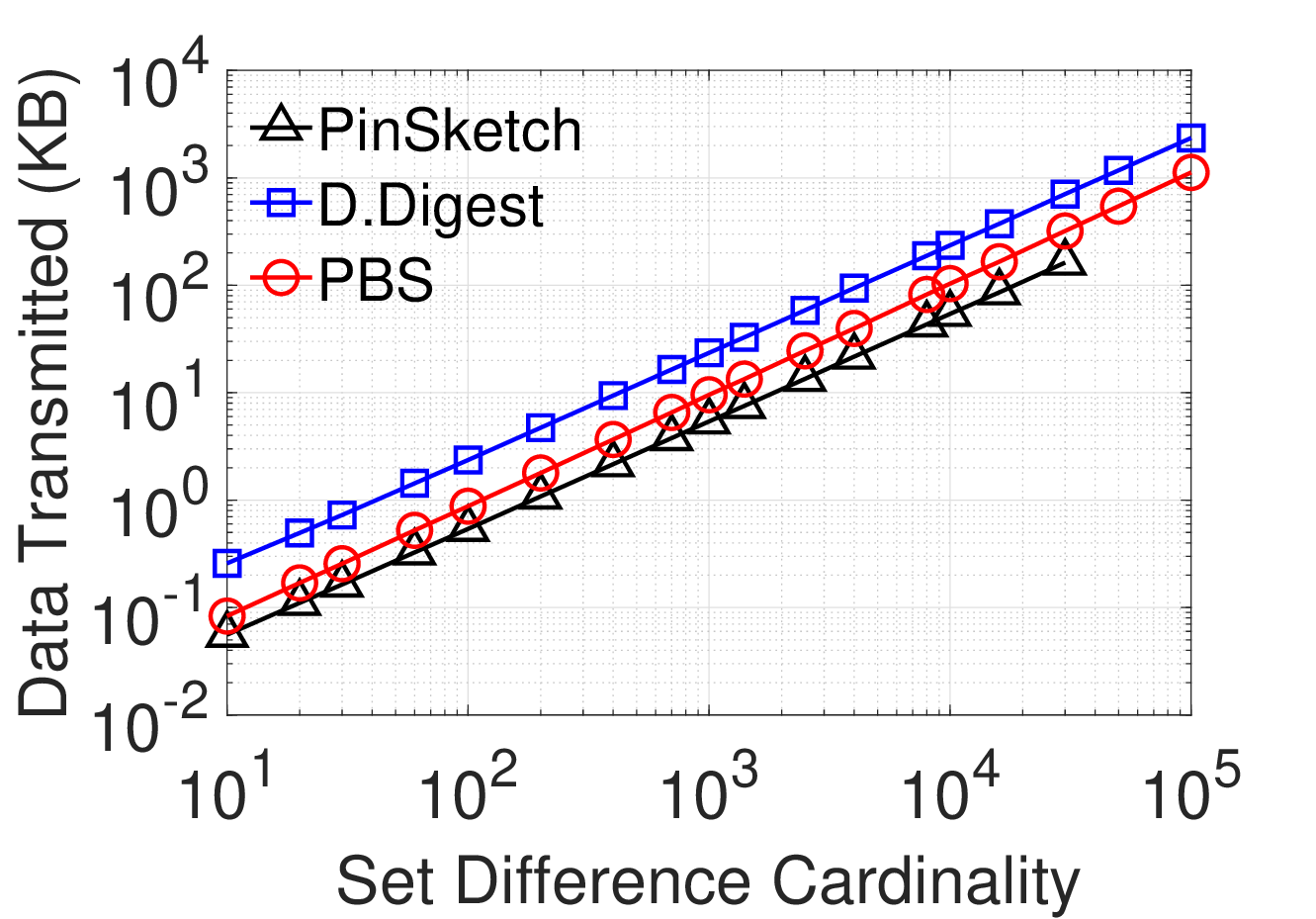}
\caption{Communication overhead}
\label{fig:co-99}
\end{subfigure}%
\begin{subfigure}[t]{0.25\textwidth}
\centering
\includegraphics[width=0.99\textwidth]{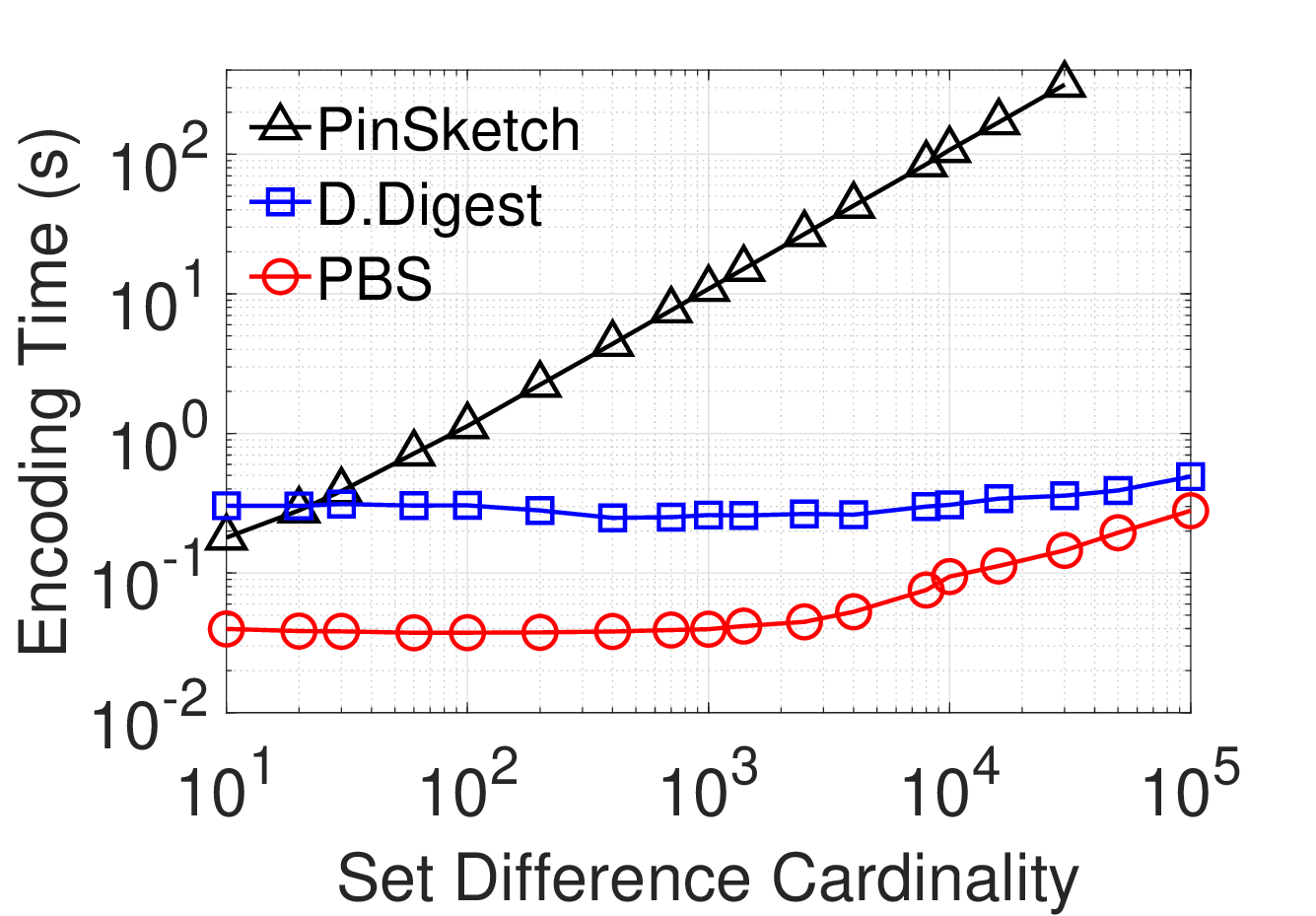}
\caption{Encoding time}
\label{fig:et-99}
\end{subfigure}%
\begin{subfigure}[t]{0.25\textwidth}
\centering
\includegraphics[width=0.99\textwidth]{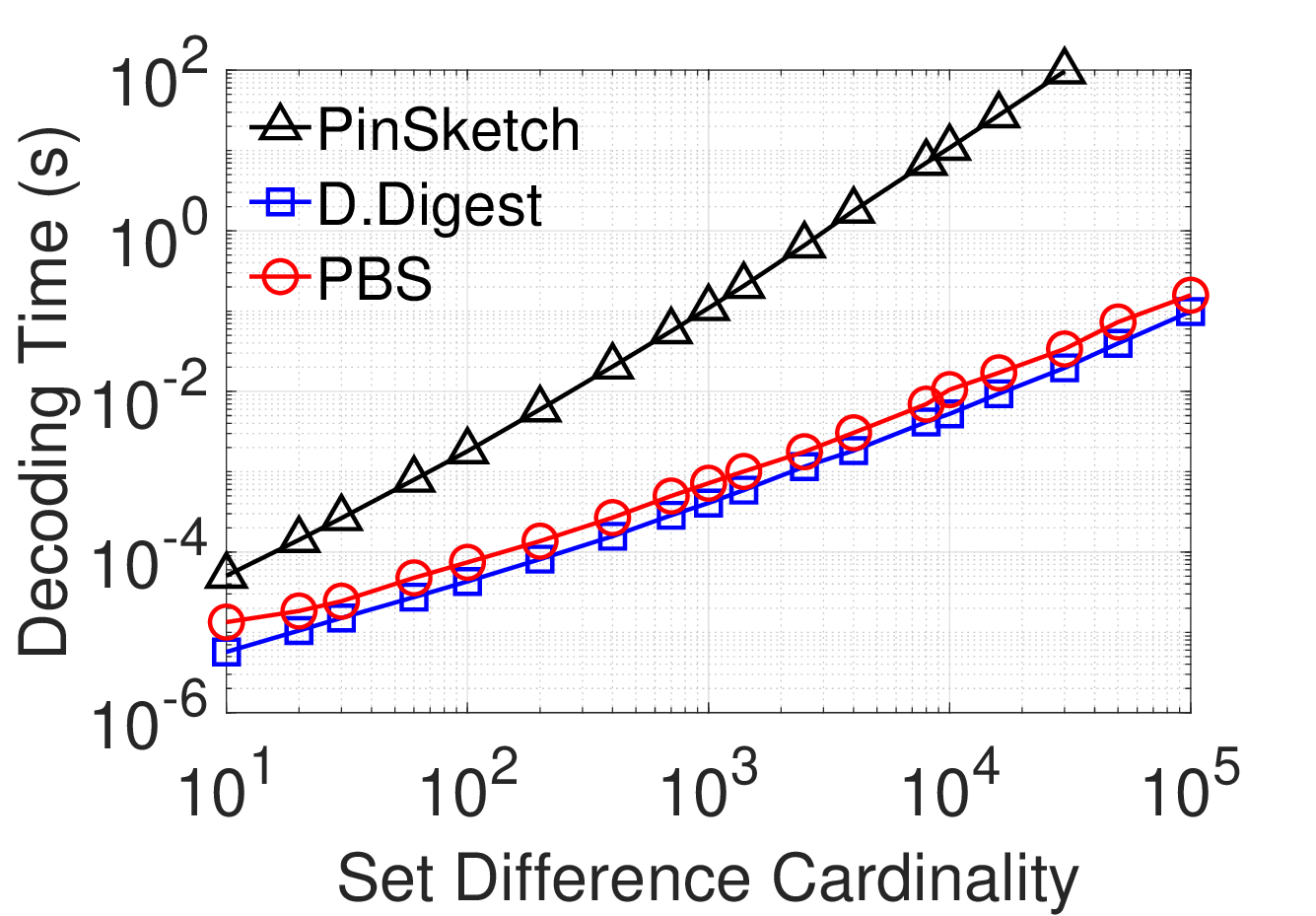}
\caption{Decoding time}
\label{fig:dt-99}
\end{subfigure}%
\caption{Comparisons against PinSketch and D.Digest, with a target success rate of 0.99.}\label{fig:comp99}
\end{minipage}\\ 
\begin{minipage}{\textwidth}
\centering
\begin{subfigure}[t]{0.25\textwidth}
\centering
\includegraphics[width=0.99\textwidth]{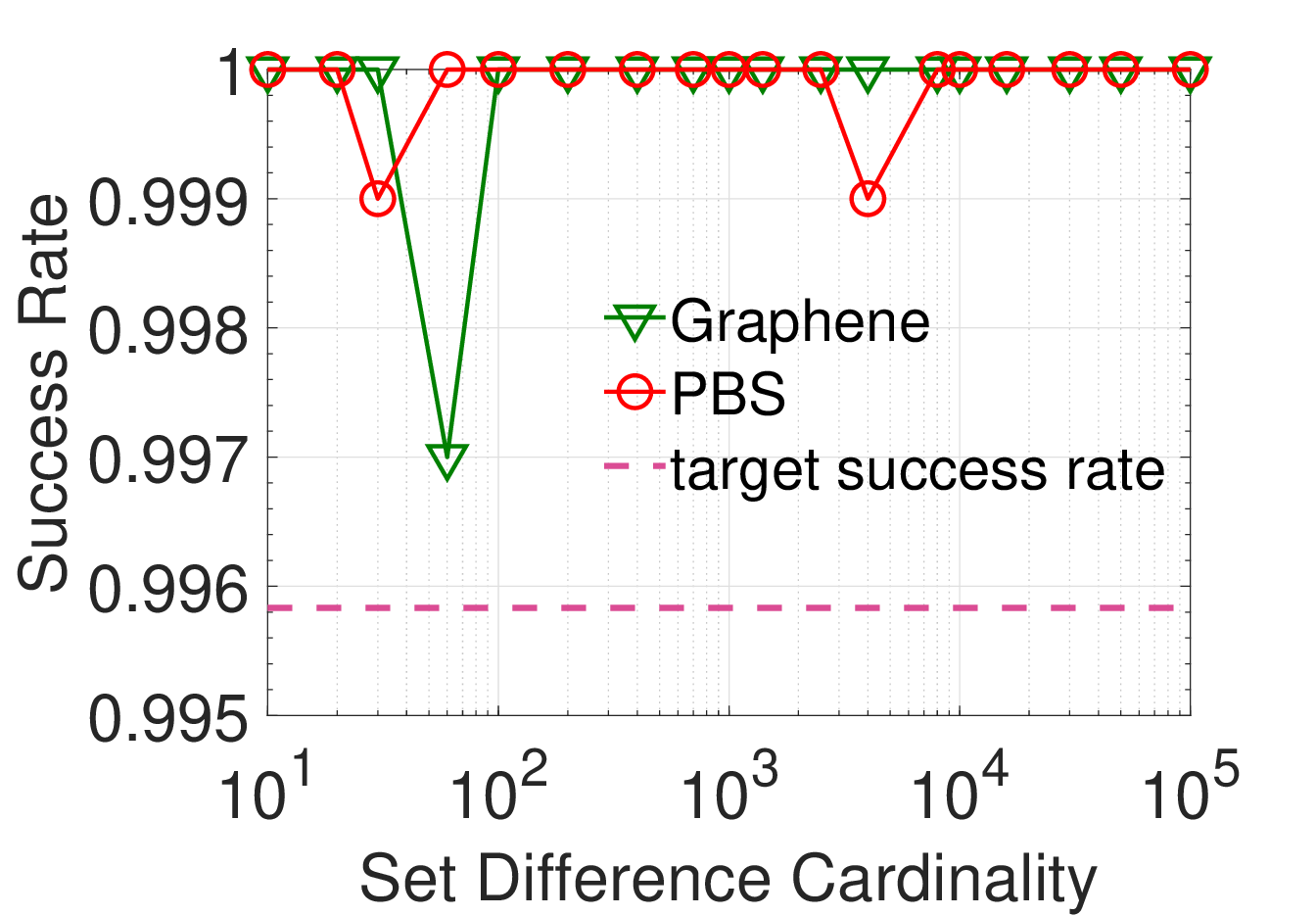}
\caption{Success rate}
\label{fig:success-rate-9958}
\end{subfigure}%
\begin{subfigure}[t]{0.25\textwidth}
\centering
\includegraphics[width=0.99\textwidth]{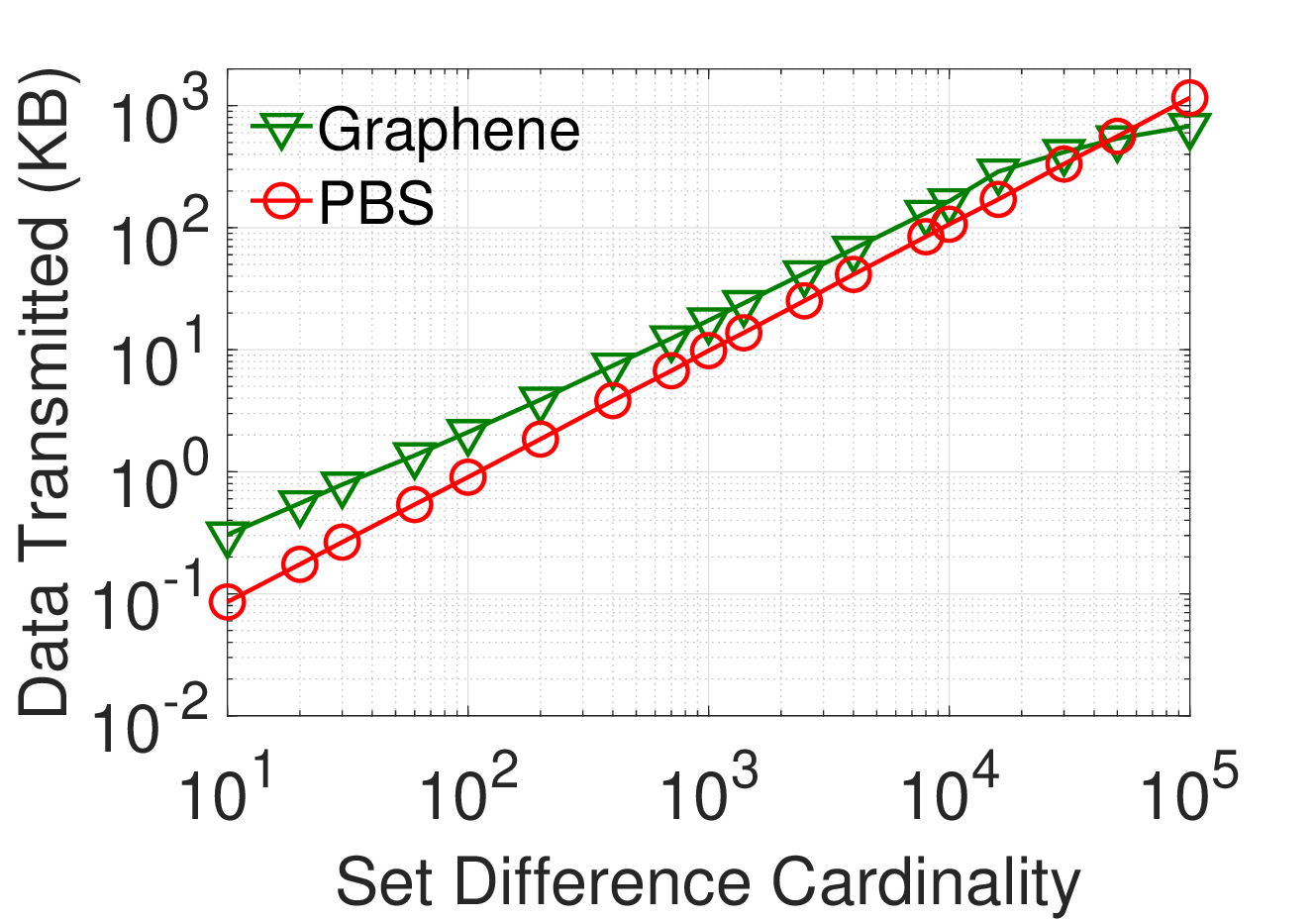}
\caption{Communication overhead}
\label{fig:co-9958}
\end{subfigure}%
\begin{subfigure}[t]{0.25\textwidth}
\centering
\includegraphics[width=0.99\textwidth]{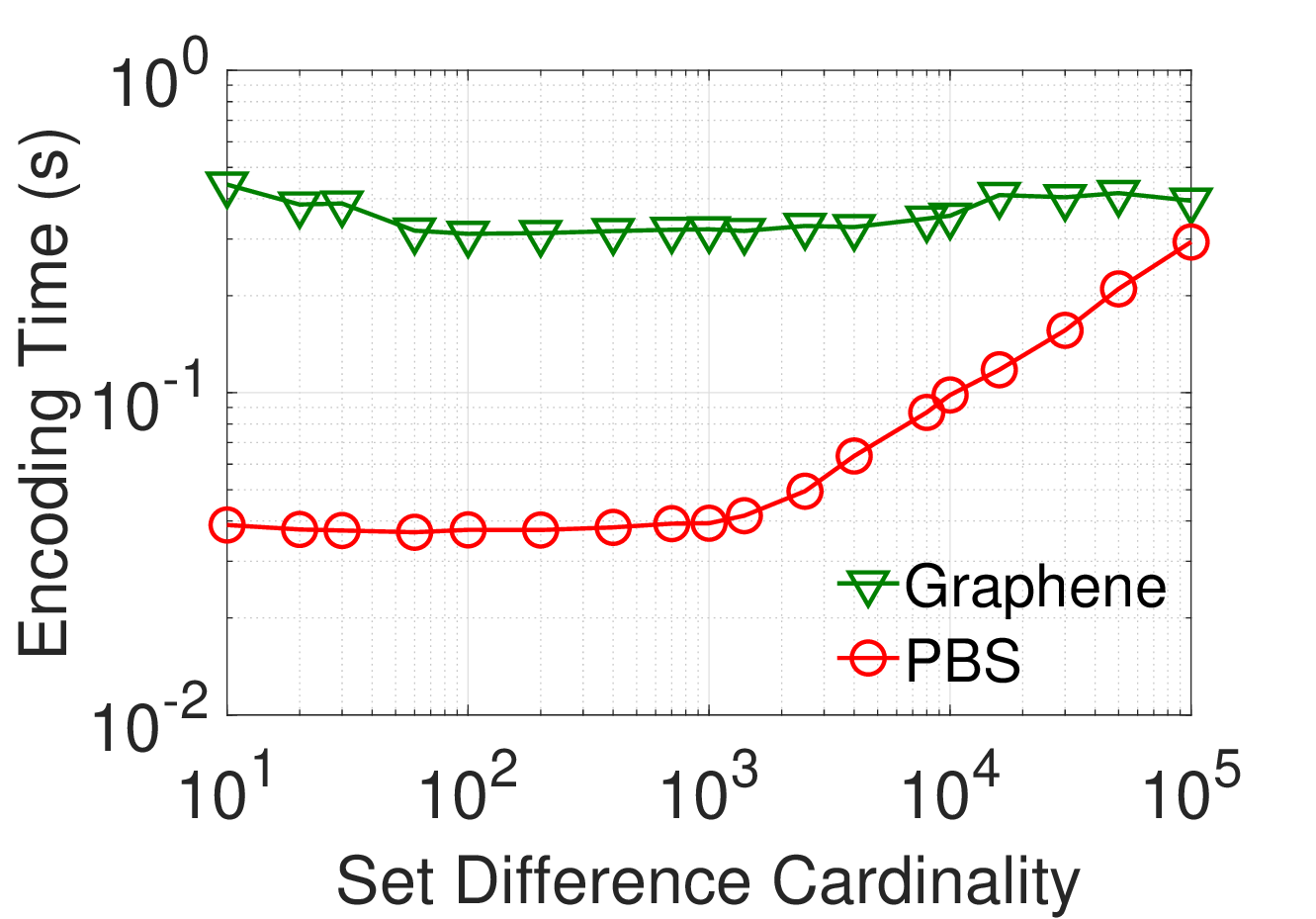}
\caption{Encoding time}
\label{fig:et-9958}
\end{subfigure}%
\begin{subfigure}[t]{0.25\textwidth}
\centering
\includegraphics[width=0.99\textwidth]{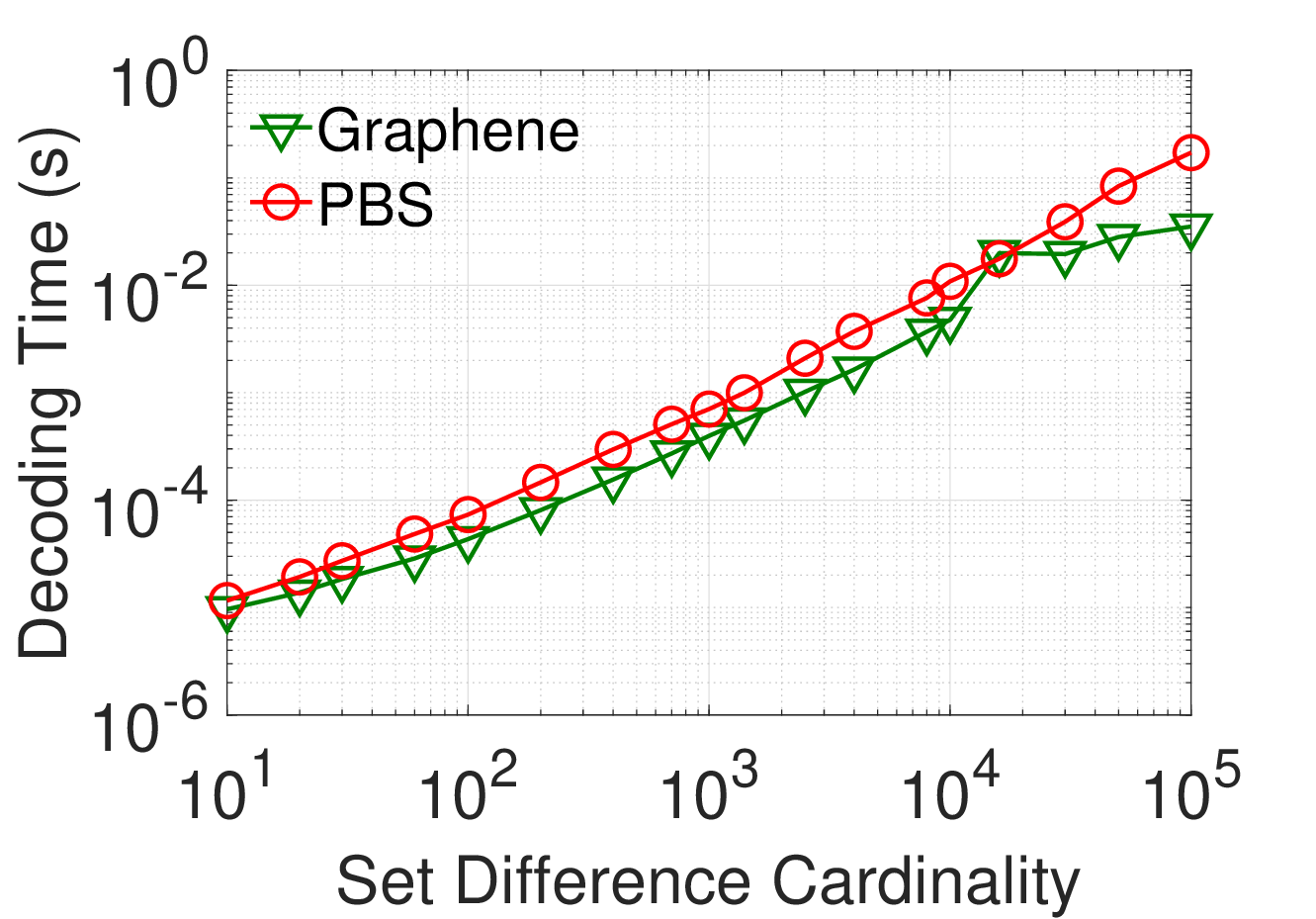}
\caption{Decoding time}
\label{fig:dt-9958}
\end{subfigure}%
\caption{Comparisons against Graphene, with a target success rate of 239/240.}\label{fig:comp9958}
\end{minipage}\\ 
\begin{minipage}{\textwidth}
\centering
\begin{subfigure}[t]{0.25\textwidth}
\centering
\includegraphics[width=0.99\textwidth]{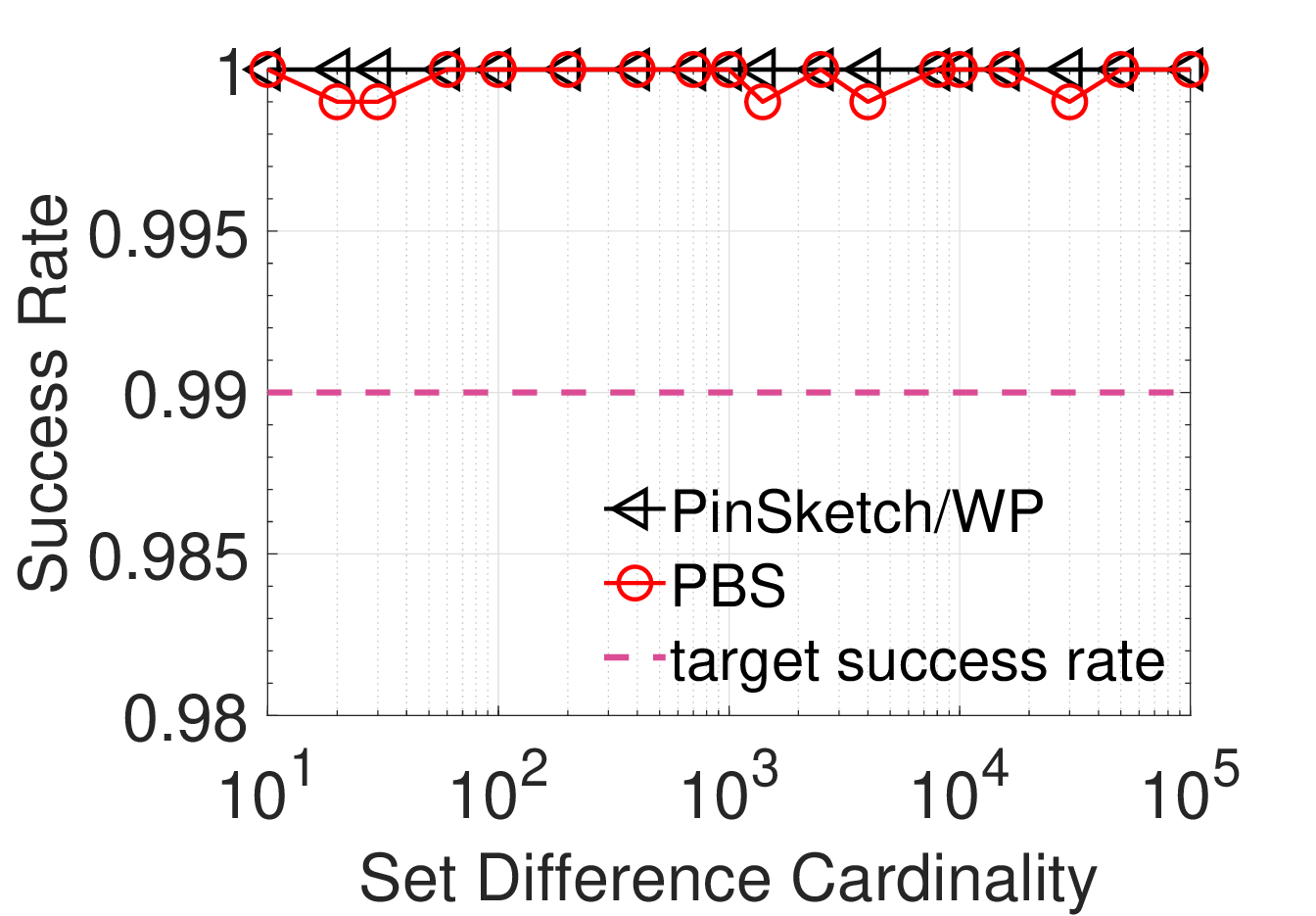}
\caption{Success rate}
\label{fig:success-rate-99-pswp}
\end{subfigure}%
\begin{subfigure}[t]{0.25\textwidth}
\centering
\includegraphics[width=0.99\textwidth]{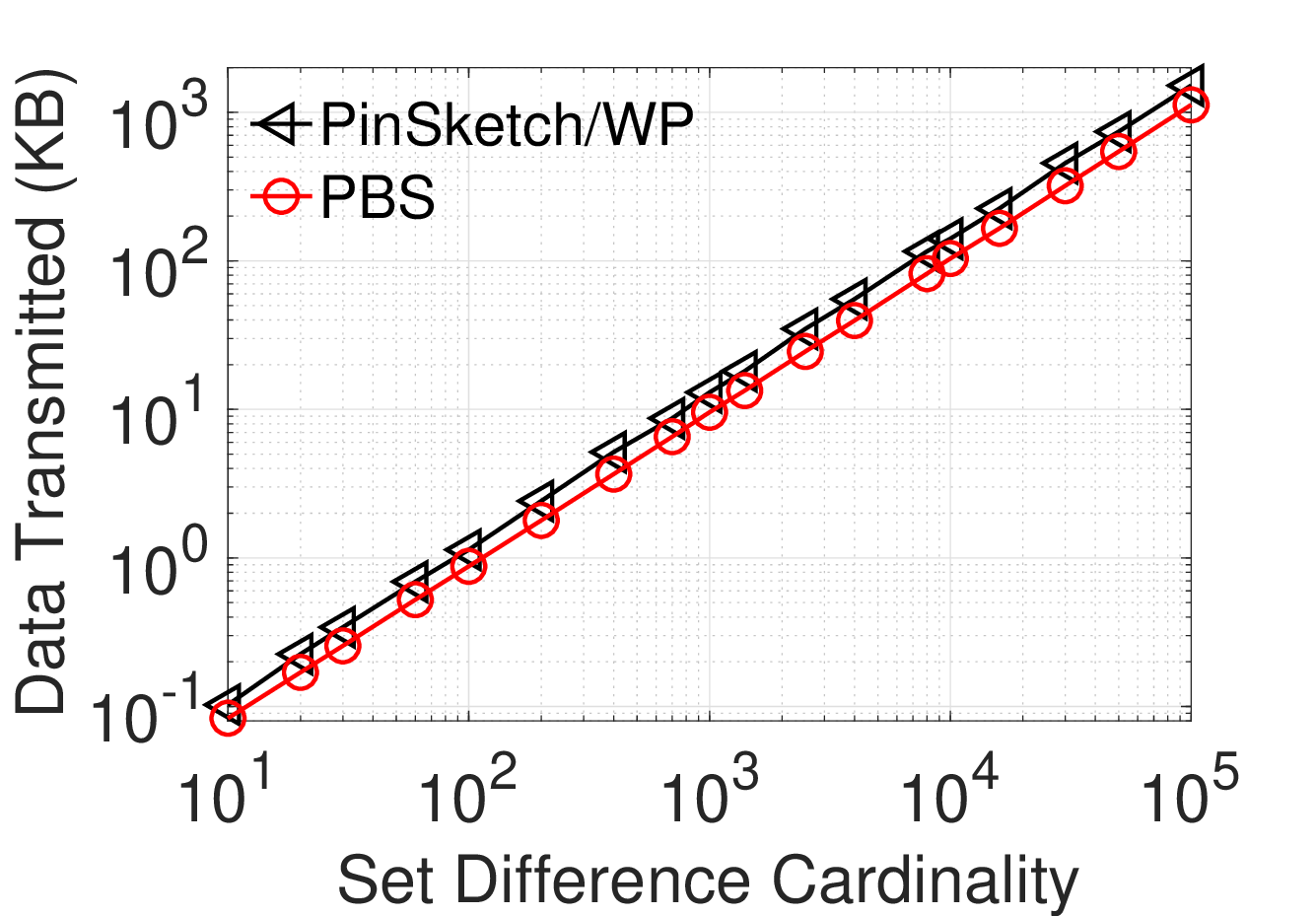}
\caption{Communication overhead}
\label{fig:co-99-pswp}
\end{subfigure}%
\begin{subfigure}[t]{0.25\textwidth}
\centering
\includegraphics[width=0.99\textwidth]{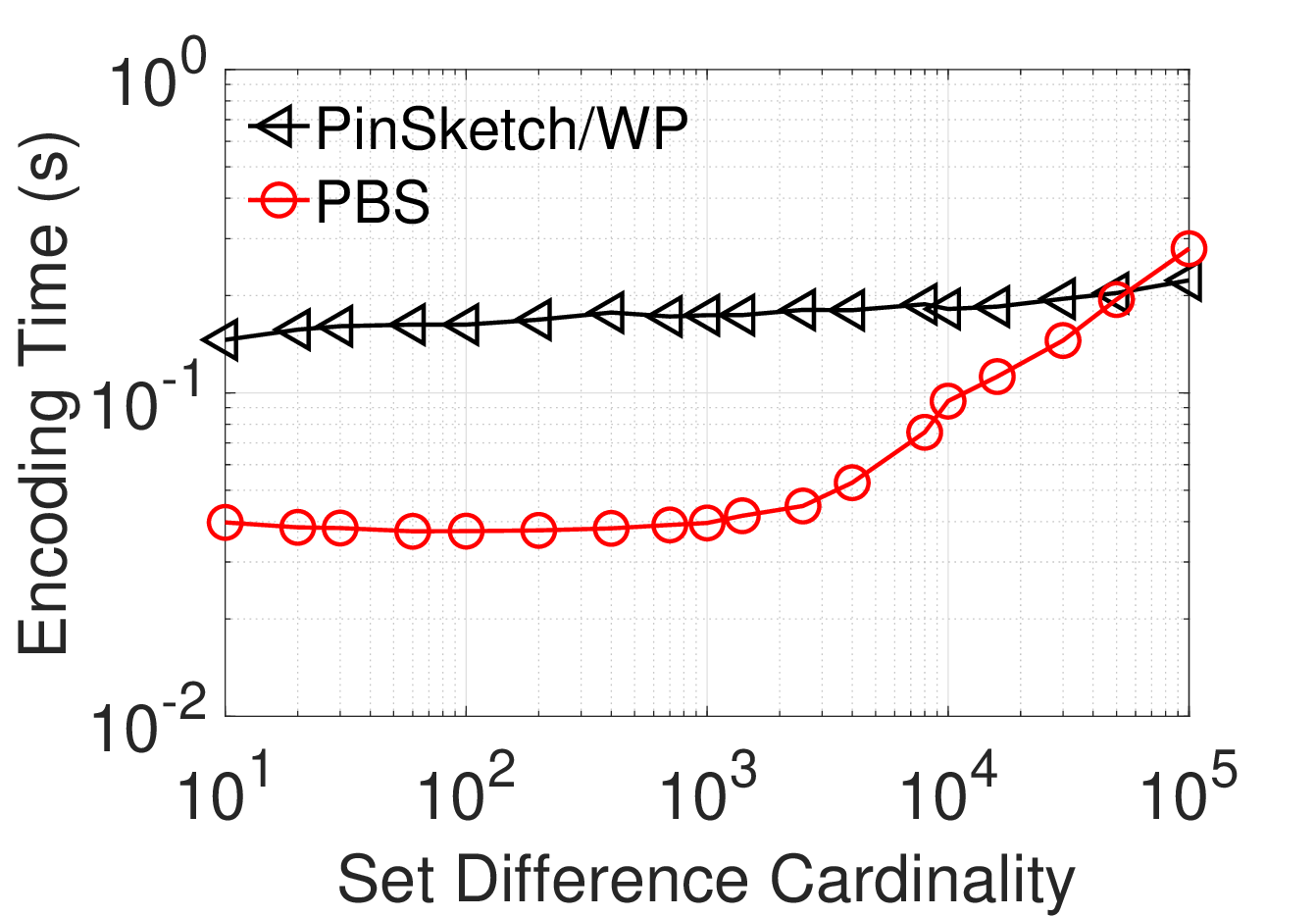}
\caption{Encoding time}
\label{fig:et-99-psbwp}
\end{subfigure}%
\begin{subfigure}[t]{0.25\textwidth}
\centering
\includegraphics[width=0.99\textwidth]{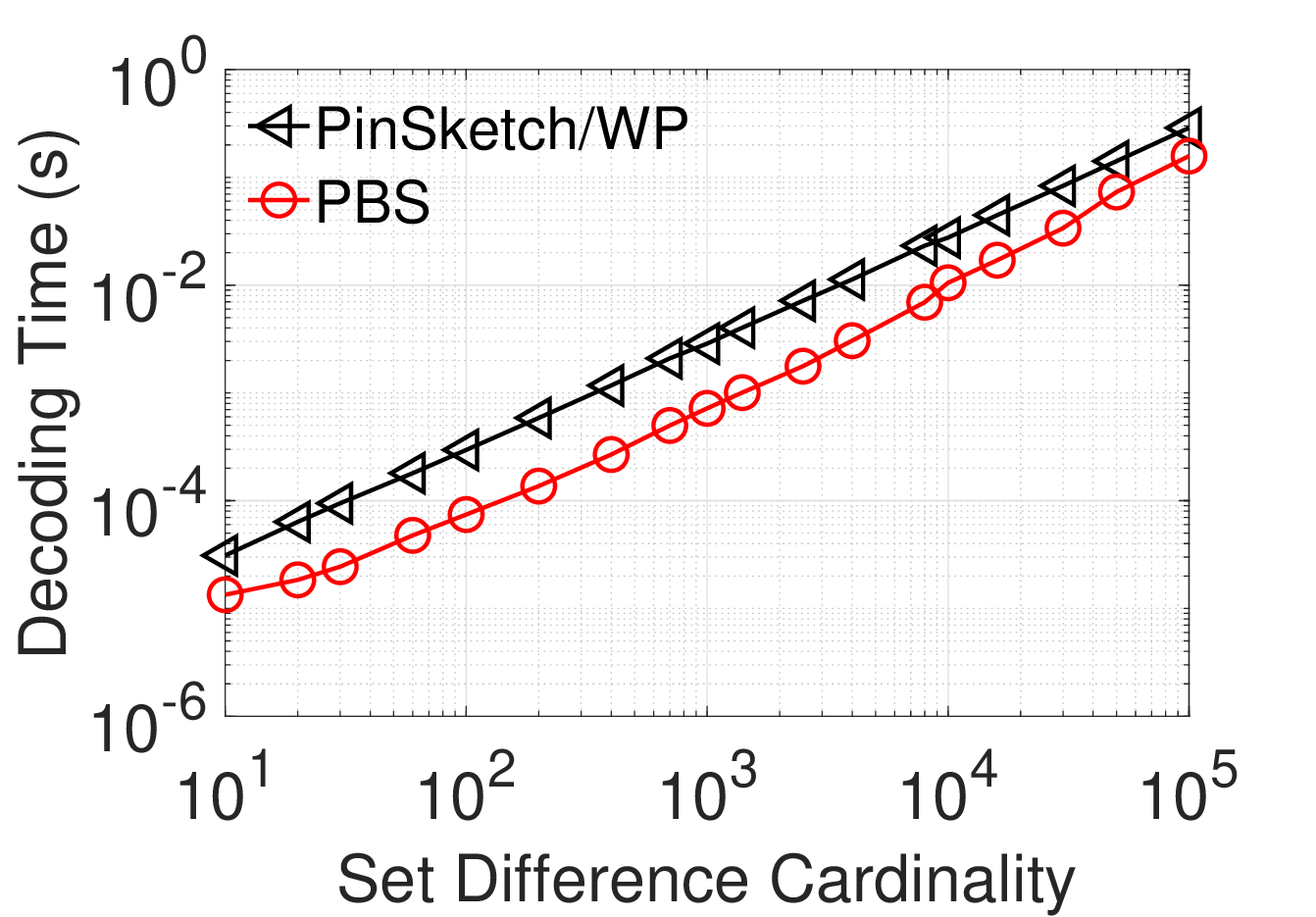}
\caption{Decoding time}
\label{fig:dt-99-pbswp}
\end{subfigure}%
\caption{Comparisons against PinSketch/WP, with a target success rate of 0.99.}\label{fig:comp99-pswp}
\end{minipage}
\end{figure*}

\subsection{PBS vs. PinSketch and D.Digest}\label{subsec:vs-ddigest-pinsketch}

In this section, we compare PBS with PinSketch and D.Digest. 
We keep the comparison of PBS with Graphene separate in~\autoref{subsec:vs-graphene}, because 
a fair comparison there calls for slightly different experimental settings and parameters. 

\subsubsection{Parameter configurations}\label{subsubsec:configuration}
 
As explained earlier, in virtually all applications, a set reconciliation algorithm should guarantee 
a high enough {\it success rate (probability)} of reconciling all distinct elements in $A\Delta B$, 
and guaranteeing a higher success rate 
generally requires higher communication and computational overheads. 
Hence, to fairly compare these set reconciliation algorithms, we should properly configure their 
parameters so that they roughly have the same success rate. 
In~\cite{Eppstein_WhatsDifference?Efficient_2011}, the authors 
have provided configuration guidelines for tuning D.Digest to achieve a success rate of 0.99. 
To tune the parameters of D.Digest to achieve other success rates, however, requires a large number of Monte-Carlo experiments.   
Instead, we tune the parameters of PinSketch and PBS to match this success rate of D.Digest, because 
it is much easier to do so for PinSketch (to be shown next) and 
PBS (shown in~\autoref{sec:parameter_settings}).

\smallskip
\noindent
{\bf PinSketch}. As explained earlier in~\autoref{subsec:ToW}, 
$Pr[d \le 1.38\hat{d}]\!\ge\!0.99$, when $\hat{d}$ is obtained from the ToW estimator with 128 sketches. 
We set the BCH error-correction capacity $t$
to $1.38\hat{d}$ so that 
the event $\{d\!\le\! t\}$ 
which corresponds to successful BCH decoding and hence set reconciliation, has a probability of at least 0.99. 

\smallskip
\noindent
{\bf D.Digest}.  
As suggested in~\cite{Eppstein_WhatsDifference?Efficient_2011}, we use $2\hat{d}$ cells 
(to both account for the randomness of $\hat{d}$ and allow accurate IBF decoding) 
in the IBF of D.Digest, 
and use $3$ hash functions if $\hat{d}$ is greater than $200$ and 4 hash functions otherwise.

\smallskip
\noindent
{\bf PBS}.  We choose $r\!=\!3$ rounds since it is a sweet spot as explained in~\autoref{sec:vary_r}. 
We set the target success probability $p_0$ to 0.99 
and optimally parameterize PBS using the procedure described in~\autoref{sec:parameter_settings}. 
{\color{red}In each experiment, we allow PBS to run at most $3$ rounds and its communication and computational 
overheads are measured as the total during all rounds executed.
}

\subsubsection{Experimental results} 

We report the experimental results in this section. 
{\color{red}
Note that we were not able obtain the results for
PinSketch for $d>\num{30000}$ in a reasonable amount of time,
as it is prohibitively expensive computationally to
do so. 
}
We first present the 
success rates of all three algorithms 
in~\autoref{fig:success-rate-99}, which shows that all these algorithms have achieved success rates 
at least as high as the target success rate (0.99) under all 
different values of $d$ except for D.Digest whose success rates are slightly lower than 0.99  
when $d\!\le\!30$. 

\smallskip
\noindent
{\bf Communication Overhead}.
\autoref{fig:co-99} compares the communication overhead of 
PBS against those of PinSketch and D.Digest. 
Results show that the communication overheads scale approximately linearly with respect to $d$ for all 
three algorithms. More precisely, for any $d$, the amount of communication per 
distinct element is roughly a constant. 
D.Digest is the worst. It requires around 6$\times$32 bits per distinct element, 
6 times the theoretical minimum (32 bits per distinct element). 
{\color{red}PBS is much better, the communication overhead of which 
is between 2.13 to 2.87 times the theoretical minimum.} 
PinSketch 
has the lowest communication overhead, 
which is 1.38 times the theoretical minimum.

\smallskip
\noindent
{\bf Encoding Time}.
\autoref{fig:et-99} compares the encoding time of PBS against those of PinSketch and D.Digest. 
\autoref{fig:et-99} clearly shows that the former  
is much lower than the latter under all different values of $d$. 

\smallskip
\noindent
{\bf Decoding Time}.
\autoref{fig:dt-99} compares the decoding time of PBS against those of PinSketch and D.Digest.
As shown in~\autoref{fig:dt-99}, the decoding time of 
PinSketch is much higher than those of D.Digest and PBS when $d$ is large (say $\ge$ \num{1000}). 
For example, when $d\!=\!$\num{10000}, the decoding time of PinSketch is roughly three orders of magnitude 
higher than those of D.Digest and PBS. 
{\color{red}\autoref{fig:dt-99} also shows clearly that D.Digest is the best, whose decoding time is $1.53$ to $2.35$ times 
shorter than that of PBS.} 

However, as discussed earlier, the encoding time of D.Digest is much (up to one order of magnitude) longer 
than that of PBS, and encoding time (of 
PBS and D.Digest) is usually much longer than the corresponding decoding time. 
Thus, PBS has the lowest overall computational overhead.

\subsection{PBS vs. Graphene}\label{subsec:vs-graphene}

In this section, we compare PBS against Graphene. 
Recall that in our experimental setting, we have $B\subset A$ and need to let Alice learn $A\Delta B$, which 
was shown in~\cite{Ozisik2019} to be the best-case scenario for Graphene in terms of communication overhead and decoding time.
Hence we have treated Graphene more than fairly here.  
Since the parameters in the source code provided by the authors of~\cite{Ozisik2019} are already optimized for 
achieving a target success rate of 239/240, we tune PBS to match this success rate:  As shown in~\autoref{fig:success-rate-9958}, 
the success rates of both PBS and Graphene are higher than 239/240. 

\smallskip
\noindent
{\bf Communication Overhead}.
\autoref{fig:co-9958} 
compares the communication overhead of PBS against that of Graphene.  
{\color{red}It shows that, even in this best-case scenario for Graphene,
PBS has much lower (roughly 1.2 to 7.4 times less) communication overhead than Graphene 
under all different values of $d$ except when $d$ gets very close to \num{100000}.
The reason behind this exception was explained earlier (in~\autoref{sec:related}):  When $d$ is sufficiently large with 
respect to $|A|$ ($=\! 10^6$ in this case), it becomes more communication-efficient overall for Graphene to start using 
a BF to reduce the size of its IBF.  
It can be calculated using an optimization formula in~\cite{Ozisik2019} that the breakeven point (for using a BF) 
in this case is some number between $d\!=\!\num{10000}$ and $d\!=\!\num{16000}$.  We can actually see in~\autoref{fig:co-9958} that the slope of the Graphene curve, which 
corresponds to the average communication 
overhead per distinct element, starts to decrease after the breakeven point, resulting in it eventually going under the PBS curve roughly after $d \!\ge\! \num{50000}$.}

{\color{red}
\smallskip
\noindent
{\bf Encoding Time}.
\autoref{fig:et-9958} clearly shows that the encoding time of PBS  
is $1.34$ to $11.38$ times lower than that of Graphene under all values of $d$. 

%
\smallskip
\noindent
{\bf Decoding Time}.
\autoref{fig:dt-9958} compares the decoding time of PBS against those of Graphene and 
clearly shows that the former is slightly ($1.20$ to $2.28$ times) longer than the latter except 
when $d$ is close to $\num{100000}$ where the former is up to $4.87$ times longer. 
}

\subsection{PBS vs. PinSketch with Partition}\label{subsec:vs-pinsketchwp}


{\color{red}
Arguably, the same algorithmic trick ({\it i.e.}, hash-partition $A$ and $B$ each into groups) can 
be applied also to PinSketch~\cite{Dodis_PinSketch_2008} 
for reducing its BCH decoding time from $O(d^2)$ to $O(d)$.  
Doing so however makes the communication overhead of PinSketch higher than that of PBS for the following reason.
As explained in~\autoref{subsec:parameterize-n-t}, 
we need to leave a safety margin in setting the BCH error-correction capacity $t$, in the sense that $t$ needs to be ``comfortably'' larger than $\delta$, the average number
of ``bit errors'' per group.   
Hence for each group pair, the average additional communication overhead (incurred for transmitting a longer BCH codeword) of leaving this safety margin
is $(t-\delta)\log n$ in PBS and is $(t-\delta)\log |\mathcal{U}|$ in PinSketch.
However, as explained in \autoref{sec:related}, 
$\log n$ is typically $3$ to $4$ times smaller than $\log |\mathcal{U}|$.  
Hence PinSketch pays $3$ to $4$ times more for leaving the safety margin, resulting in a higher overall communication overhead, as we elaborate next.  

Now we compare the performance of PBS against that of PinSketch with hash partition, which we refer to as PinSKetch/WP. 
For PinSketch/WP, we use the 
same $\delta$ and $t$ values as in PBS (there is no parameter $n$ in PinSketch/WP since it does not use a parity bitmap), 
with a target success probability of $p_0$=0.99 within $r$=3 rounds, in each experiment instance. 
The experimental results are reported in~\autoref{fig:comp99-pswp}.  It clearly shows that 
PBS outperforms PinSketch/WP in both communication 
overhead and computational overhead (the sum of the encoding and the decoding time).  
Note this outperformance will increase when the hash signature length $\log|\mathcal{U}|$ increases ($\log|\mathcal{U}| \!=\! 32$ 
bits in~\autoref{fig:comp99-pswp}).
Hence, PBS would outperform by a wider margin in real-world blockchain applications 
where $\log|\mathcal{U}|$ is much larger (e.g., $\log|\mathcal{U}| \!=\! 256$ bits in Bitcoin~\cite{Nakamoto_Bitcoinpeerpeer_2019}), 
as has been shown in~\refToAppendix{\autoref{app-sec:vs-pinsketchwp}}{Appendix J.3 in~\cite{Gong_pbs-tr_2020}}.

}

\section{Conclusion}\label{sec:conclusion}

In this paper, we propose Parity Bitmap Sketch (PBS), a space- and computationally-efficient solution to the set reconciliation problem. 
We show, through experiments, that PBS strikes a much better tradeoff between communication and 
computational overheads than all the state-of-the-art solutions. 
In addition, we derive a novel rigorous analytical framework for PBS, which most existing solutions do not have. 
Through three applications of this framework, we demonstrate that it enables both 
the accurate analysis of various performance metrics such as success probability and the tuning of the parameters of PBS for near-optimal performances.  

\balance
\bibliographystyle{abbrv}
\bibliography{bib/NTGfull,bib/set_difference,bib/added_during_revision}  

\appendix

\section{Mean and Variance Proof for ToW Estimator}\label{app-sec:mean-var-proof}

{\color{red}
We now prove the claims for the mean and variance of the ToW estimator we made in~\autoref{subsec:tow-estimator}. 
Note that, the proof for the mean is almost the same as the proof in~\cite{AlonMS99}; that for the variance is 
almost the same as that in~\cite{Hua_simplerbetterdesign_2012}. We reproduce them with some minor changes (to 
adapt them to our context) for this paper to be self-contained. 

In the following, we first state a well-known fact concerning the four-wise independent hash family that stems from its 
definition~\cite{Wegman_k-wise-independent_1979}. 
\begin{fact}\label{fact:4-wise}
If a hash family $\mathcal{F}=\{f: \mathcal{U}\rightarrow \mathcal{Y}\}$ is four-wide independent, then for any distinct 
$s_1,s_2,s_3,s_4\in\mathcal{U}$, the hash values $f(s_1),f(s_2),f(s_3),f(s_4)$ are independent and identically distributed random variables, each of which is uniformly distributed in $\mathcal{Y}$.
\end{fact}

We now proceed to prove that the ToW estimator is unbiased, {\it i.e.,} $\mathbf{E}[\hat{d}]=d$, as follows.
\begingroup
\allowdisplaybreaks
\begin{eqnarray}
\mathbf{E}[\hat{d}] &=& \mathbf{E}\left[\left(Y_f(A) - Y_f(B)\right)^2\right] \notag \\
          &=& \mathbf{E}\left[\left(\sum_{a \in A} f(a) - \sum_{b \in B} f(b)\right)^2\right] \notag \\
          &=& \mathbf{E}\left[\left(\sum_{s \in A\setminus B} f(s) - \sum_{s \in B\setminus A} f(s)\right)^2\right] \notag \\
          &=& \mathbf{E}\left[\sum_{s \in A\setminus B} f^2(s) + \sum_{s_1,s_2\in A\setminus B\land s_1\neq s_2}f(s_1)f(s_2)  \right. \notag\\
          &&- 2\sum_{s_1\in A\setminus B\land s_2\in B\setminus A}f(s_1)f(s_2) \notag\\
          &&+ \sum_{s_1,s_2\in B\setminus A\land s_1\neq s_2}f(s_1)f(s_2)  \notag\\
          &&+ \left. \sum_{s \in B\setminus A} f^2(s)\right]  \notag \\
          &=& \mathbf{E}\left[\sum_{s \in A\setminus B} f^2(s)\right] + E\left[\sum_{s_1,s_2\in A\setminus B\land s_1\neq s_2}f(s_1)f(s_2)\right] \notag \\
          &&- 2\cdot\mathbf{E}\left[\sum_{s_1\in A\setminus B\land s_2\in B\setminus A}f(s_1)f(s_2)\right] \notag\\
          &&+ \mathbf{E}\left[\sum_{s_1,s_2\in B\setminus A\land s_1\neq s_2}f(s_1)f(s_2)\right]  \notag\\
          &&+ \mathbf{E}\left[\sum_{s \in B\setminus A} f^2(s)\right]  \notag \\
          &=& |A\setminus B| + 0 - 0 + 0 + |B\setminus A| \label{eqn:mean-expl} \\
          &=& |A\triangle B| = d \notag
\end{eqnarray}
\endgroup

Equation~\autoref{eqn:mean-expl} holds for the following reasons. 
For any $s\in \mathcal{U}$, we have $f^2(s)=1$ because $\mathcal{Y}=\{1,-1\}$. Hence, 
the first and last terms on the LHS of~\autoref{eqn:mean-expl} are equal to the corresponding terms on the RHS of~\autoref{eqn:mean-expl}, respectively. For the second term on the LHS of~\autoref{eqn:mean-expl}, it is equal to $0$ because of~\autoref{fact:4-wise} and the 
linearity of expectation. For the same reason, each of the rest two terms on the LHS of~\autoref{eqn:mean-expl} is also equal to $0$. 

Now, we prove that the claim we made for the variance of $\hat{d}$, {\it i.e.,} $\mbox{\bf{Var}}[\hat{d}]=2d^2-2d$, as follows. 
\begin{eqnarray}
\mathbf{E}[\hat{d}^2] &=& \mathbf{E}\left[\left(Y_f(A) - Y_f(B)\right)^4\right] \notag \\
  &=& \mathbf{E}\left[\left(\sum_{a \in A} f(a) - \sum_{b \in B} f(b)\right)^4\right] \notag \\
  &=& \mathbf{E}\left[\sum_{s \in A\triangle B} f^4(s)\right] \notag \\
  &&+ \mathbf{E}\left[ 3 \sum_{s_1,s_2\in A\triangle B\land s_1 \neq s_2 } f^2(s_1)f^2(s_2)\right]   \label{eqn:second-moment-expl}\\
  &=& d + 3d(d - 1) \notag\\
  &=& 3d^2 - 2d \notag
\end{eqnarray}

We note that no terms in the form of $f(s_1)f^3(s_2)$ (for any distinct $s_1,s_2$) appear on the RHS of~\autoref{eqn:second-moment-expl}, because any such term equals $f(s_1)f(s_2)$, the expectation of which, 
as explained earlier, is equal to $0$.

Using the above second moment and the following expression for the variance, of $\hat{d}$, we have, 
\begin{equation}
    \mbox{\bf{Var}}[\hat{d}] = \mathbf{E}[\hat{d}^2] - \left(\mathbf{E}[\hat{d}]\right)^2 = 2d^2 - 2d
\end{equation}
}

\section{Other Estimators}\label{subsec:other-estimators}

Past set reconciliation work mostly uses two other estimators: the min-wise estimator and 
the Strata estimator~\cite{Eppstein_WhatsDifference?Efficient_2011}.
The min-wise estimator is derived from min-wise hashing technique~\cite{Broder_MinWiseIndependent_2000} 
for estimating set 
Jaccard similarity $J(A,B)\!\triangleq\! |A\cap B|/|A \cup B |$. 
The Strata estimator is 
based on 
the idea of Flajolet and Martin (FM) sketch~\cite{Flajolet_Strata_1985}, originally designed for estimating $F_0$, 
the zeroth frequency moment of a data stream.   
In comparison, 
the ToW estimator is much more space-efficient 
according to our experiments under various parameter settings.  
The results are not shown here in the interest of space. 

The sketch, proposed in~\cite{Feigenbaum_AppL1Difference_2003} and designed for 
estimating the $L^1$-difference between two functions, can also be 
used to estimate the set difference cardinality. It has similar space-efficiency as the ToW estimator. 
However, it is much less computationally-efficient, as it requires a nontrivial amount of computation for
constructing random variables that are 
range-summable -- a property that is not needed for our application. 

\section{Correctness of PBS}\label{app-sec:correctness}
{\color{red}
In this section, we formally prove the correctness of PBS, more precisely PBS-for-large-d. 
It suffices to prove the correctness of PBS-for-small-d, because the reconciliation of 
each group pair in PBS-for-large-d is solved by PBS-for-small-d. 
The following theorem states the correctness of PBS-for-small-d.

\begin{theorem}\label{thm:corrctness}
Assume that the checksum verification step in~\autoref{step:terminate_condition} 
of~\autoref{alg:single_round_pbs} produces correct verifications (in every round).  If PBS-for-small-d terminates after $r$ rounds, then 
\[
\hat{D}_1\triangle\hat{D}_2\triangle\cdots\triangle \hat{D}_r=A\triangle B.
\]
\end{theorem}
\begin{proof}
Since PBS-for-small-d terminates after $r$ rounds, the checksum verification step (\autoref{step:terminate_condition} of~\autoref{alg:single_round_pbs}) must have succeeded in the $r^{th}$ round.
Hence we have
$c\big{(}(A\triangle\hat{D}_1\triangle\hat{D}_2\triangle\cdots\triangle \hat{D}_{r-1})\triangle \hat{D}_{r}\big{)}=c(B)$. 
Since we assume that the checksum verification step produces correct verifications, we have
$(A\triangle\hat{D}_1\triangle\hat{D}_2\triangle\cdots\triangle \hat{D}_{r-1})\triangle \hat{D}_{r}=B$. Therefore,
\[
\hat{D}_1\triangle\hat{D}_2\triangle\cdots\triangle \hat{D}_r=A\triangle B.
\]
\end{proof}
}

\section{Preciseness of Markov-Chain Modeling}\label{sec:preciseness}

{\color{red}The Markov-chain model 
in~\autoref{sec:single_markov}
precisely captures every detail of the set reconciliation process in each group pair except for one event: 
when a fake distinct element resulting from a type (\RN{2}) exception 
is indeed not a distinct element but ``luckily'' (unluckily for Alice) 
passes the sub-universe verification check (\autoref{alg:verification}) 
and gets included in the estimated set difference $\hat{D}$ in the correspoinding round.}  
We choose to leave this event out in the modeling, because 
to incorporate it would result in a much more complicated Markov chain.   To incorporate this event is also not necessary for 
our modeling to be numerically accurate, because this event happens with a tiny probability of $O(10^{-7})$.  



In this model,
we assume $x\! \le\! t$, where $t$ is the error-correction capacity of the BCH code.
To extend the Markov-chain modeling to the case of 
$x\! >\! t$ (resulting in a BCH decoding failure w.h.p.) is a daunting task, since this group pair is to 
be partitioned into $3$ sub-group-pairs as described in \autoref{sec:BCH_exception} and each sub-group-pair is to be independently reconciled. 
We also found this task unnecessary:  Assuming $Pr[x\!\xrsquigarrow{r}\! 0]\! =\! 0$ when $x\!>\!t$ in our modeling (always to our disadvantage) 
results in only a slight underestimation of ``good'' probabilities, as will be elaborated in \autoref{sec:times2bound}.

\section{How To Compute \texorpdfstring{\larger\boldmath{$M$}}{M}}\label{sec:compute_M}

Although the transition probability matrix $M$ is in theory an infinite matrix, in this work for all practical purposes (e.g., for 
computing $Pr[x\!\xrsquigarrow{r}\! 0]$), it can be considered a $(t+1) \times (t+1)$ matrix, where $t$ is the BCH error-correction capacity.  
In this section, we describe how to compute the transition probability matrix $M$ in detail. 
{\color{red}To compute $M(i,j)$ is not straightforward for the following reason.  
Each state $j$ with $j>3$ in the Markov chain is a {\it composite state} 
consisting of a large number of 
{\it atom states}.
Only the transition probability from state $i$ to any atom state (of state $j$) can be stated as a closed-form expression
 (more precisely, a multinomial formula) and computed straightforwardly. 
The value of $M(i,j)$ is the total of all the transition probabilities from state $i$ to each of the atom states of state $j$.} 
{\color{red}Since the number of atom states grows exponentially with $j$, it is 
complicated (as it is necessary to enumerate all atom states), error-prone, and computationally expensive 
to compute $M(i, j)$ this way when $j$ is large (say $j>12$), as we will elaborate next.}

{\color{red}Each atom state of state $j$ is, in combinatorics terms, {\it a permutation of a combination of $j$}, which here corresponds to how these $j$ balls are 
distributed in the $n$ bins (by the hash function).  For a simple example, when $j = 4$ and $n = 7$, the vector $(2, 0, 0, 0, 0, 2, 0)$ is such an atom state, which corresponds to these $7$ bins (in a predefined order such as the natural order) having $2, 0, 0, 0, 0, 2,$ and $0$ ``bad'' balls in them respectively. 
Clearly, the number of such atom states (vectors) grows exponentially with $j$. For instance, when $j=13, 14, 15, 16, 17$, 
the number of distinct atom state vectors is $2.47\times 10^{12}$, $2.10\times 10^{13}$, $1.11\times 10^{14}$, $8.03\times 10^{14}$, $4.34\times 10^{15}$ respectively.  Although the computation of $M(i,j)$ can be simplified by the fact that 
two atom states have the same transition probability (i.e., belong to the same equivalence class) 
from the same state $i$ if one's state vector (e.g., $(2, 0, 0, 0, 0, 2, 0)$ in the example above)
is a permutation of the other's (e.g., $(2, 0, 2, 0, 0, 0, 0)$), grouping such a large number of atom states into equivalence classes and
computing the total transitioning probability of each is a complicated and error-prone process.  
}




{\color{red}Our solution is to decompose each composite state $j$ into a much small number of coarse-grained sub-states, each of which may still contain a large number of atom states.  Although the transition probability from a state $i$ to any sub-state of $j$ is still a summation formula and hence hard to compute
in the ``mundane'' way as explained above, we discover a recurrence relation among these transition probabilities that makes them easily computable using dynamic programming.}

We now define these sub-states and describe the recurrence relation among the resulting transition probabilities.
Each state $j$ (``bad" balls) is decomposed into $j\!+\!1$ sub-states that we denote as $(j,0)$, $(j,1)$, $(j, 2)$, $\cdots$, $(j,j)$ respectively.
Each sub-state $(j, k)$, $k = 0, 1, 2, \cdots, j$, corresponds to the set of scenarios in which the $j$ ``bad'' balls end up occupying exactly $k$ ``bad'' bins
(those containing more than one ``bad'' balls as defined earlier in \autoref{sec:single_markov}).  Let $(i, j, k)$ denotes the event that throwing $i$ balls results in
the sub-state $(j, k)$.  Let $\tilde{M}(i,j,k)$ denote the probability of this event $(i, j, k)$.  
Clearly, we have $M(i,j) = \sum_{k=0}^{j} \tilde{M}(i,j,k)$ for $\forall i,j$.



To obtain $M$, it suffices to compute the probabilities of $\tilde{M}(i,j,k)$ for $\forall i,j,k=0,1,\cdots,t$. 
To do so, we derive a recurrence relation for these probability values by rendering the process of throwing $i$ balls ``in slow motion'' in the following sense: We throw them one at a time.  Now consider what events happen before and after the $i^{th}$ ball is thrown.   Suppose the event $(i, j, k)$ happens after the $i^{th}$ ball is thrown.  There are only three possible events before the $i^{th}$ ball is thrown that can lead to the
event $(i, j, k)$.  Here we explain only the first event $(i\!-\!1,j\!-\!2,k\!-\!1)$ in detail, as explanations for the other two are similar.  
The event $(i\!-\!1,j\!-\!2,k\!-\!1)$ leads to the event $(i, j, k)$ if and only if the $i^{th}$ ball lands in a previously ``good'' bin 
so that the ``good'' ball 
already in the bin and the $i^{th}$ ball turn into two ``bad'' balls.  The latter happens with probability 
$(i\!-\!j\!+\!1)/n$, since $(i\!-\!1)\!-\!(j\!-\!2)$ out of $n$ bins are ``good'' at 
sub-state $(j\!-\!2,k\!-\!1)$.  
The other two events are $(i\!-\!1,j\!-\!1,k)$ and $(i\!-\!1,j,k)$, and they lead to the event $(i, j, k)$ with probability $k/n$ and $1\!-\!(i\!-\!1\!-\!j\!+\!k)/n$ respectively.
Summarizing these three cases, we can express $\tilde{M}(i,j,k)$, the probability of the event $(i, j, k)$, as the weighted sum of the probabilities of these three events
as follows:

\begin{align}
&    \frac{i\!-\!j\!+\!1}{n}\cdot \tilde{M}(i\!-\!1,j\!-\!2,k\!-\!1)+\frac{k}{n}\cdot \tilde{M}(i\!-\!1,j\!-\!1,k)\nonumber\\
&+(1\!-\!\frac{i\!-\!1\!-\!j\!+\!k}{n})\cdot \tilde{M}(i\!-\!1,j,k)\nonumber
\end{align}




Exploiting these recurrence relations (for $\forall i,j,k$), we can compute the values of all the $(t+1)^3$ $\tilde{M}(i, j, k)$ terms in $O(t^3)$ time via dynamic programming, for which the probability formulae of the three types of base cases are:


\begin{align}
\tilde{M}(0,j,k) =& \left\{ \begin{array}{ccl}
1, & \mbox{if }
 j=k=0 \\ 0, & \mbox{otherwise} 
\end{array}\right.\nonumber\\
\tilde{M}(i,0,k) =& \left\{ \begin{array}{ccl}
\binom{n}{i}/n^i, & \mbox{if }
 k=0 \\ 0, & \mbox{otherwise} 
\end{array}\right.\nonumber\\
\tilde{M}(i,j,0) =& \left\{ \begin{array}{ccl}
\binom{n}{i}/n^i, & \mbox{if }
 j=0 \\ 0, & \mbox{otherwise} 
\end{array}\right.\nonumber
\end{align}

\section{Success probability analysis}\label{sec:times2bound}


As explained earlier in \autoref{sec:pbs-for-large-d}, in PBS-for-large-$d$, the sets $A$ and $B$ each needs to be hash-partitioned into $g$ groups.  
Let $\delta_1, \delta_2, \cdots, \delta_g$ be the numbers of distinct elements contained in these $g$ group pairs respectively.
It is not hard to check that random variables $\delta_1, \delta_2, \cdots, \delta_g$ are identically distributed with distribution 
$Binomial(d, 1/g)$ (since each of the $d$ distinct elements between $A$ and $B$ is uniformly at random hash-partitioned into 
one of the $g$ groups).   Let $R_i$, $i=1,2,...,g$, be the number of rounds it takes for the reconciliation of group pair $i$ to be successfully completed.  
Then by definition, we have $Pr[R_i \le r] = Pr[\delta_i \!\xrsquigarrow{r}\! 0]$.  Since $\delta_1, \delta_2, \cdots, \delta_g$ have the same distribution $Binomial(d, 1/g)$, 
it suffices to derive $Pr[\delta_1 \!\xrsquigarrow{r}\! 0]$, the success probability for group pair $1$, as the success probability for any other group pair is identical.  
Replacing $x$ by $\delta_1$ in Formula \autoref{eq:single_success_prob}, 
we obtain $Pr[\delta_1 \!\xrsquigarrow{r}\! 0] = \sum_{x=0}^d Pr[\delta_1\!=\!x]\cdot Pr[x\!\xrsquigarrow{r}\! 0]$.   

As discussed in \autoref{sec:preciseness}, to our disadvantage, we assume $Pr[x\!\xrsquigarrow{r}\! 0]\!=\!0$ when $x\!>\!t$ (the BCH error-correction capacity).  
However, this assumption results in only a slight underestimation
of the success probability $Pr[\delta_1 \!\xrsquigarrow{r}\! 0]$ for the following reason.  
The Binomial term $Pr[\delta_1\!=\!x]$, the ``coefficient'' of $Pr[x\!\xrsquigarrow{r}\! 0]$, is very small when $x\!>\!t$ since $E[\delta_1]\!=\!\delta$ 
and $t$ is typically set to between $1.5$ to $3.5$ times of $\delta$.   For notational succinctness, we denote this slightly underestimated success probability 
$\sum_{x=0}^t Pr[\delta_1\!=\!x]\cdot 
Pr[x\!\xrsquigarrow{r}\! 0]$ as $\alpha$ in the sequel.   Clearly, the success probability for any other group pair is also bounded by $\alpha$.

By definition, we have $R = \max(R_1, R_2, ..., R_g)$.  However, we cannot state rigorously that the overall (for all $g$ group pairs) success probability $Pr[R\leq r]$
is lower bounded by $\alpha^g$, because each $R_i$ is a function of the random variable $\delta_i$, and the $g$ random variables $\delta_1$, $\delta_2$, ..., $\delta_g$
are not mutually independent as their sum is $d$.   However, in this special case of identically binomially distributed random variables with a fixed sum,  
we can prove, using Corollary 5.11 in \cite{upfal2005probability}, a rigorous and only slightly weaker lower bound of $1\!-\!2(1\!-\!\alpha^g)$ for $Pr[R\leq r]$:  It only doubles  
the failure probability $1 - \alpha^g$ in the unrigorous lower bound $\alpha^g$ (written as $1\!-\!(1\!-\!\alpha^g)$) to $2(1\!-\!\alpha^g)$ in 
the rigorous bound.  

\section{Analysis on ``Piecewise Reconciliability''}\label{app-sec:analysis-piecewise}

{\color{red}Again we focus our attention on the first group pair that have $\delta_1$ (distributed as $Binomial(d, 1/g)$ as explained earlier) distinct elements between them.}
Let $Z_k$, $k=1,2,...$, be the number of distinct elements among those $\delta_1$ that are reconciled in the $k^{th}$ round. 
Clearly, our goal is to compute $\mathbf{E}[Z_1]$, $\mathbf{E}[Z_2]$, $\mathbf{E}[Z_3]$, ..., and so on. 
To do so, it suffices to compute the unconditional expectations $\mathbf{E}[Z_1+Z_2+\cdots+Z_k]$ for $k=1, 2, \cdots$.
They in turn can be derived from the following conditional expectations on the LHS of Equation \autoref{eq:piecewise}.
Equation \autoref{eq:piecewise} holds because both sides calculate the expected number of distinct elements that are reconciled 
within $k$ rounds, conditioned upon the event $\{\delta_1 = x\}$.





\begin{equation}\label{eq:piecewise}
\mathbf{E}[Z_1+Z_2+\cdots+Z_k|\delta_1=x]=\sum_{y=0}^x (x-y)\cdot Pr[x\!\xrsquigarrow{k}\! y]
\end{equation}

\begin{figure*}
\centering
\begin{subfigure}[t]{0.25\textwidth}
\centering
\includegraphics[width=0.99\textwidth]{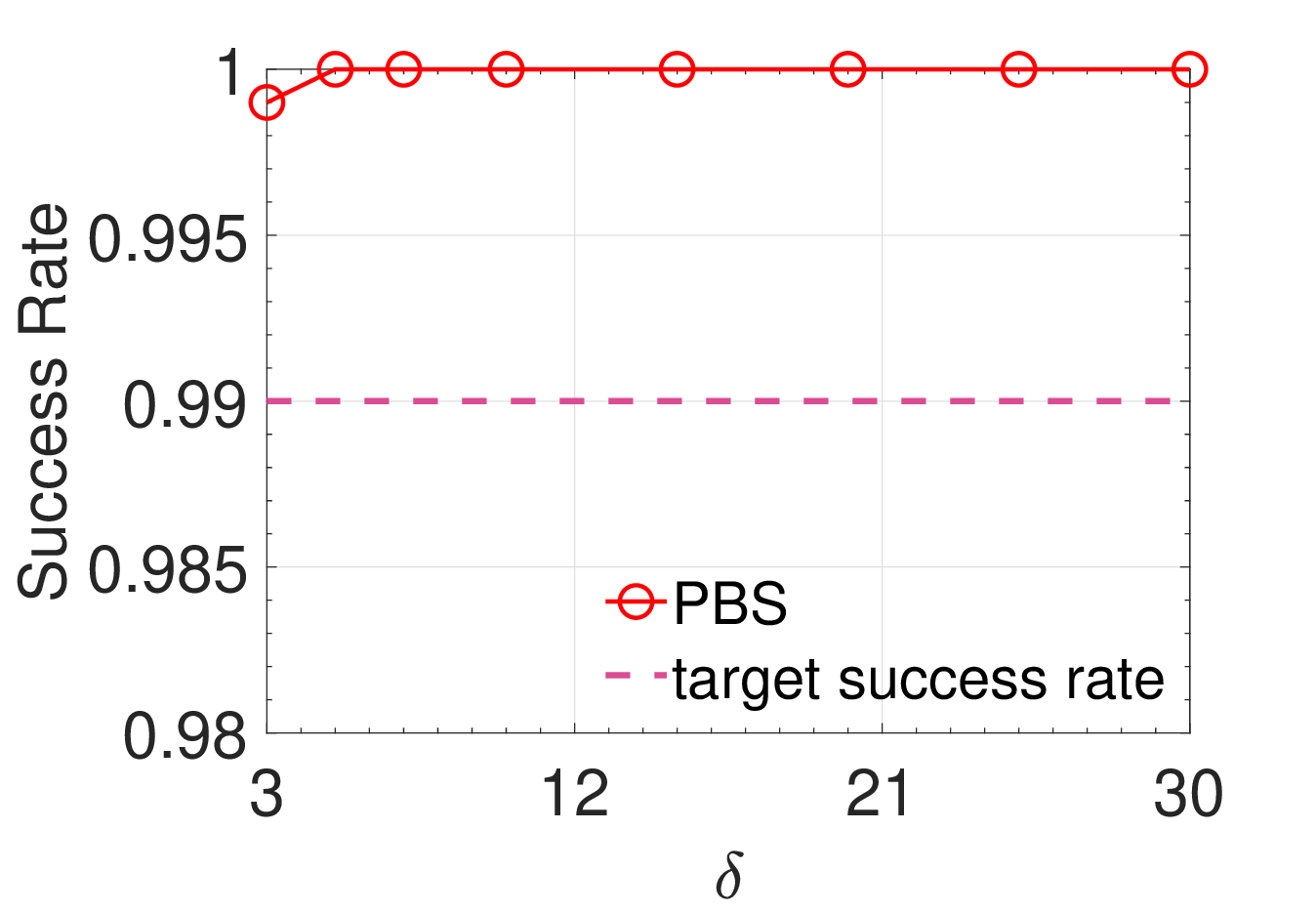}
\caption{Success rate}
\label{fig:success-rate-99-diff-avg}
\end{subfigure}%
\begin{subfigure}[t]{0.25\textwidth}
\centering
\includegraphics[width=0.99\textwidth]{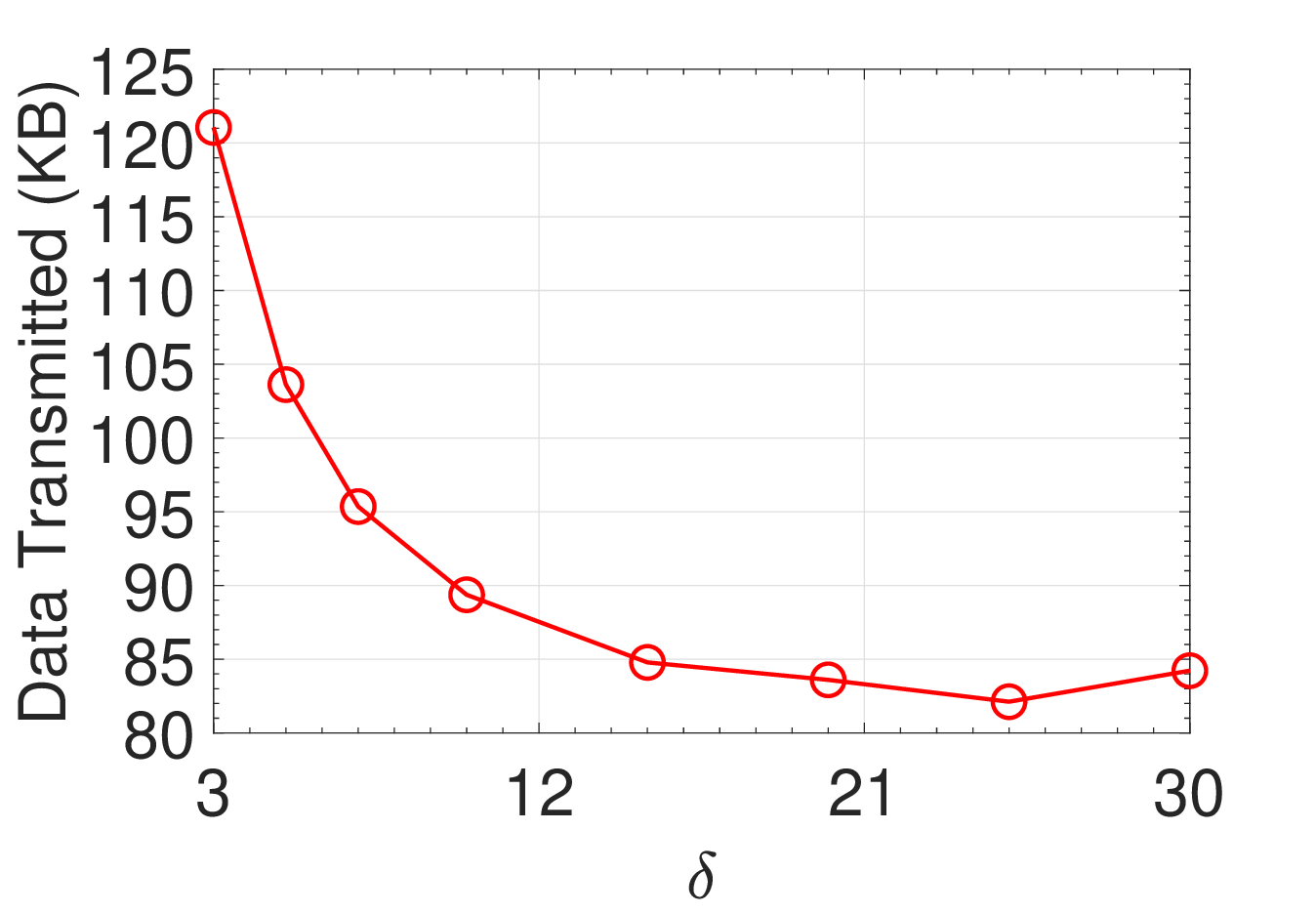}
\caption{Communication overhead}
\label{fig:co-99-diff-avg}
\end{subfigure}%
\begin{subfigure}[t]{0.25\textwidth}
\centering
\includegraphics[width=0.99\textwidth]{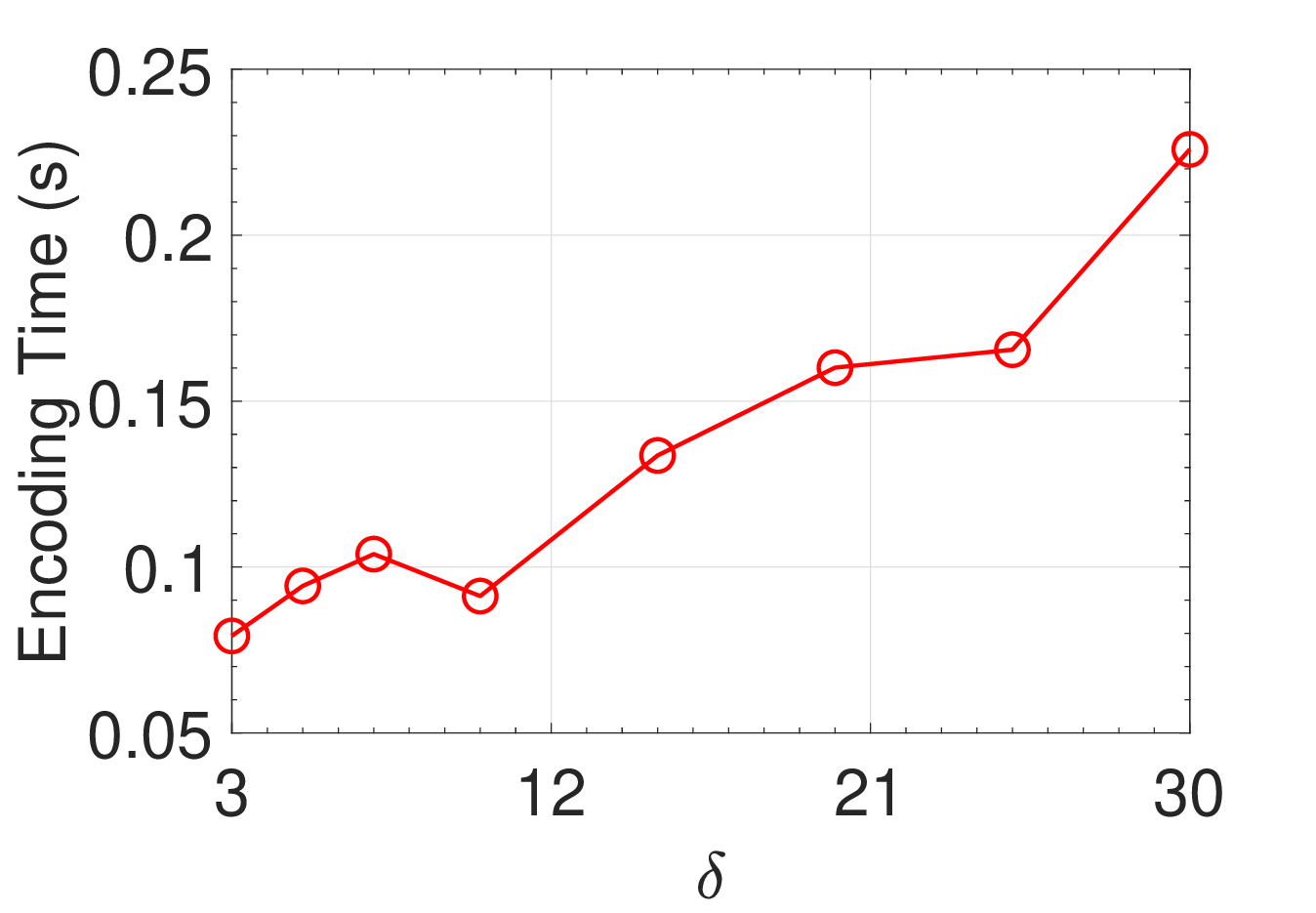}
\caption{Encoding time}
\label{fig:et-99-diff-avgs}
\end{subfigure}%
\begin{subfigure}[t]{0.25\textwidth}
\centering
\includegraphics[width=0.99\textwidth]{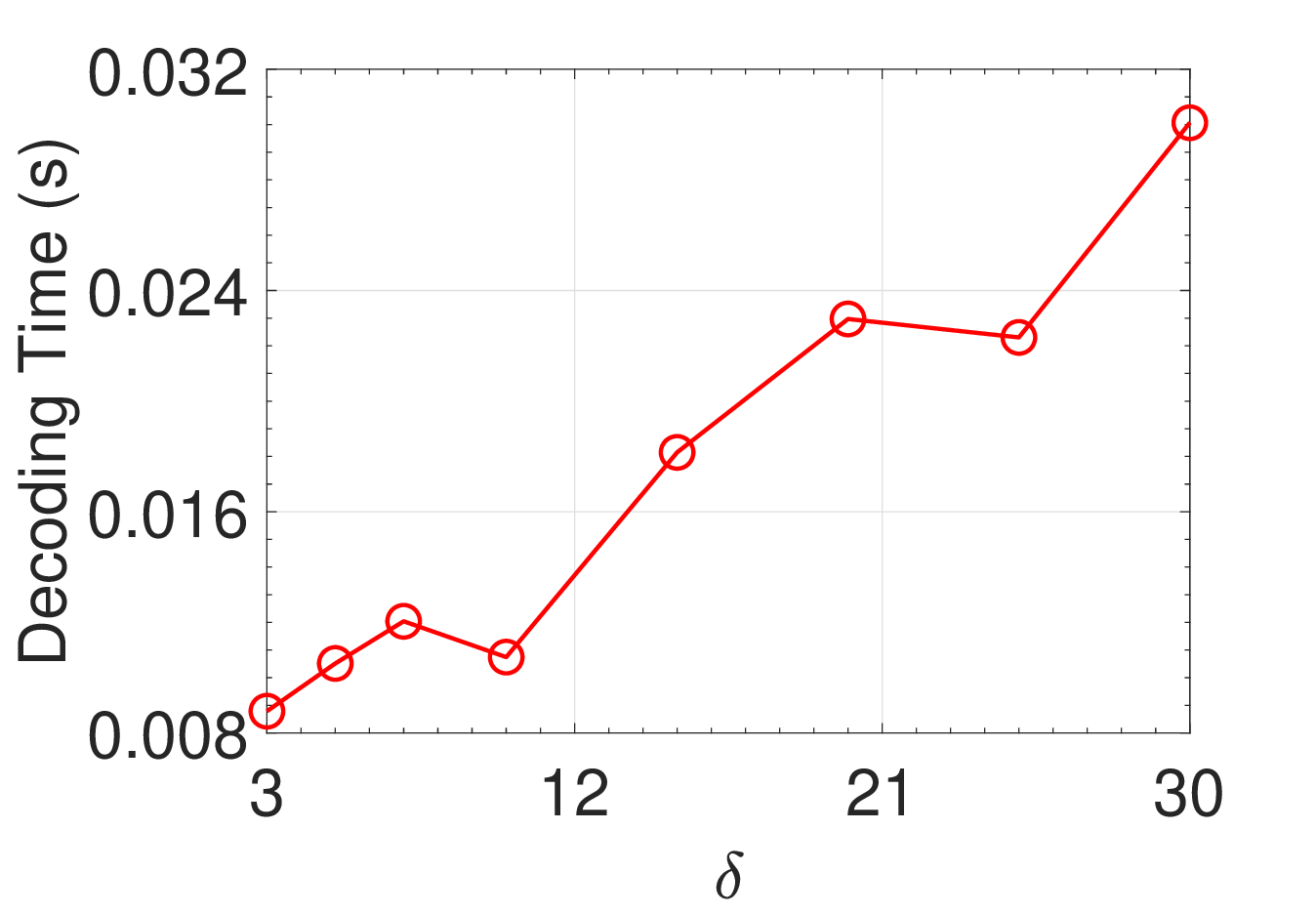}
\caption{Decoding time}
\label{fig:dt-99-diff-avgs}
\end{subfigure}%
\caption{Comparisons of PBS using different values of $\delta$, with a target success rate of 0.99.}\label{fig:comp99-diff-avgs}
\end{figure*}

\section{An Example for Parameter Optimization}\label{app:exam-param-op}

\begin{table}[!ht]
\setlength{\tabcolsep}{3.2pt}
\centering
\caption{Success probability lower bound values.}\label{tab:success_prob}
\begin{tabular}{@{}|c|c|c|c|c|c|c|@{}}
\hline 
\diagbox{$t$}{$n$}    & $63$ & $127$ & $255$ & $511$ & $1023$ & $2047$ \\ \hline 
8   & 0 & 25.5\% & 32.7\% & 34.3\% & 34.9\% & 35.0\% \\ \hline
9   & 52.1\% & 78.0\% & 84.2\% & 85.7\% & 86.1\% & 86.2\% \\ \hline
10  & 75.1\% & 92.7\% & 96.5\% & 97.4\% & 97.6\% & 97.7\% \\ \hline
11  & 85.9\% & 96.9\% & \cellcolor{blue!25}99.1\% & \cellcolor{blue!25}99.5\% & \cellcolor{blue!25}99.6\% & \cellcolor{blue!25}99.6\% \\ \hline
12  & 91.3\% & 98.5\% & \cellcolor{blue!25}99.7\% & \cellcolor{blue!25}99.9\% & \cellcolor{blue!25}$>$99.9\% & \cellcolor{blue!25}$>$99.9\% \\ \hline
13  & 93.9\% & \cellcolor{blue!60}99.1\% & \cellcolor{blue!25}99.8\% & \cellcolor{blue!25}$>$99.9\% & \cellcolor{blue!25}$>$99.9\% & \cellcolor{blue!25}$>$99.9\% \\ \hline
14  & 95.1\% & \cellcolor{blue!25}99.4\% & \cellcolor{blue!25}$>$99.9\% & \cellcolor{blue!25}$>$99.9\% & \cellcolor{blue!25}$>$99.9\% & \cellcolor{blue!25}$>$99.9\% \\ \hline
15  & 95.6\% & \cellcolor{blue!25}99.5\% & \cellcolor{blue!25}$>$99.9\% & \cellcolor{blue!25}$>$99.9\% & \cellcolor{blue!25}$>$99.9\% & \cellcolor{blue!25}$>$99.9\% \\ \hline
16  & 95.7\% & \cellcolor{blue!25}99.6\% & \cellcolor{blue!25}$>$99.9\% & \cellcolor{blue!25}$>$99.9\% & \cellcolor{blue!25}$>$99.9\% & \cellcolor{blue!25}$>$99.9\% \\ \hline
17  & 95.8\% & \cellcolor{blue!25}99.6\% & \cellcolor{blue!25}$>$99.9\% & \cellcolor{blue!25}$>$99.9\% & \cellcolor{blue!25}$>$99.9\% & \cellcolor{blue!25}$>$99.9\% \\ \hline
\end{tabular}
\end{table}

{\color{red}In this section, we illustrate our parameter optimization procedure described in~\autoref{sec:parameter_settings} with an 
example.} 
Suppose we have $d\!=\!\num{1000}$ distinct elements, $\delta\!=\!5$ (so that $g\!=\!200$ groups), $r\!=\!3$ rounds, and
target success probability $p_0\!=\!99\%$.  For each $(n, t)$ value combination in 
$\{63,127,255,511,1023,2047\}\times\{8,9,\cdots,16,17\}$
we compute the corresponding lower bound ($1-2(1-\alpha^g(n,t))$) value.  The lower bound values corresponding to these $(n, t)$ value combinations are shown \autoref{tab:success_prob}.
In \autoref{tab:success_prob}, each cell in which the corresponding lower bound value is no smaller than the target success probability $p_0\!=\!99\%$ is highlighted.
Among the highlighted cells, the cell further darkened 
results in the smallest objective function value and hence its ``coordinates" $n\!=\!127, t\!=\!13$ are the optimal 
parameter setting in this instance.  Using pre-computation, the success probability value in each cell can be computed in $O(1)$ time, so this optimization procedure is very efficient computationally.  

\section{Use of BCH in Communication}\label{app:bch}

{\color{red}In this section, we describe the standard BCH encoding 
for its usual application of communication over a noisy channel and explain how it differs from the BCH encoding in PBS.   
In the standard BCH encoding, a coded message, which is the uncoded message concatenated with the codeword, is $n = 2^m - 1$ bits long 
in total.  For the codeword to correct up to $t$ bit errors, that may occur to both the uncoded message part and the codeword part 
during the transmission of the coded message over the noisy channel, it needs to be $tm$ bits long, resulting in a ``leftover" of 
at most $n - tm$ bits for the uncoded message.

In PBS, the codeword $\xi_A$ is also $tm$ bits long, but the uncoded message $A[1..n]$ (which is not transmitted at all) can be $n$ bits 
(instead of $n - mt$ bits) long.  The uncoded message is allowed to be longer here because, unlike in the usual application 
of communication over a noisy channel where bit errors can happen also to the codeword, 
in PBS no bit error will happen to the codeword $\xi_A$ during its transmission.} 

\section{More Experimental Results}\label{app:more-experimental-results}

\subsection{Number of Rounds Required by PBS}\label{app:completed-rounds}
\begin{table}[!ht]
    \caption{Empirical probability mass function table for the number of rounds required by 
PBS to correctly reconcile all distinct elements.} 
    \label{tab:completed-rounds}
    \centering
\begin{tabular}{@{}|r | r | r | r|@{}}
\hline 
\diagbox{$d$}{$r$} &     1 &     2 &     3 \\ \hline 
$          10$ & 0.804 & 0.188 & 0.008 \\ \hline 
$         100$ & 0.217 & 0.760 & 0.023 \\ \hline 
$  \num{1000}$ & 0     & 0.957 & 0.043 \\ \hline 
$ \num{10000}$ & 0     & 0.907 & 0.093 \\ \hline 
$\num{100000}$ & 0     & 0.818 & 0.182 \\
\hline
\end{tabular}
\end{table}

\begin{figure}
\centering
\includegraphics[width=0.9\columnwidth]{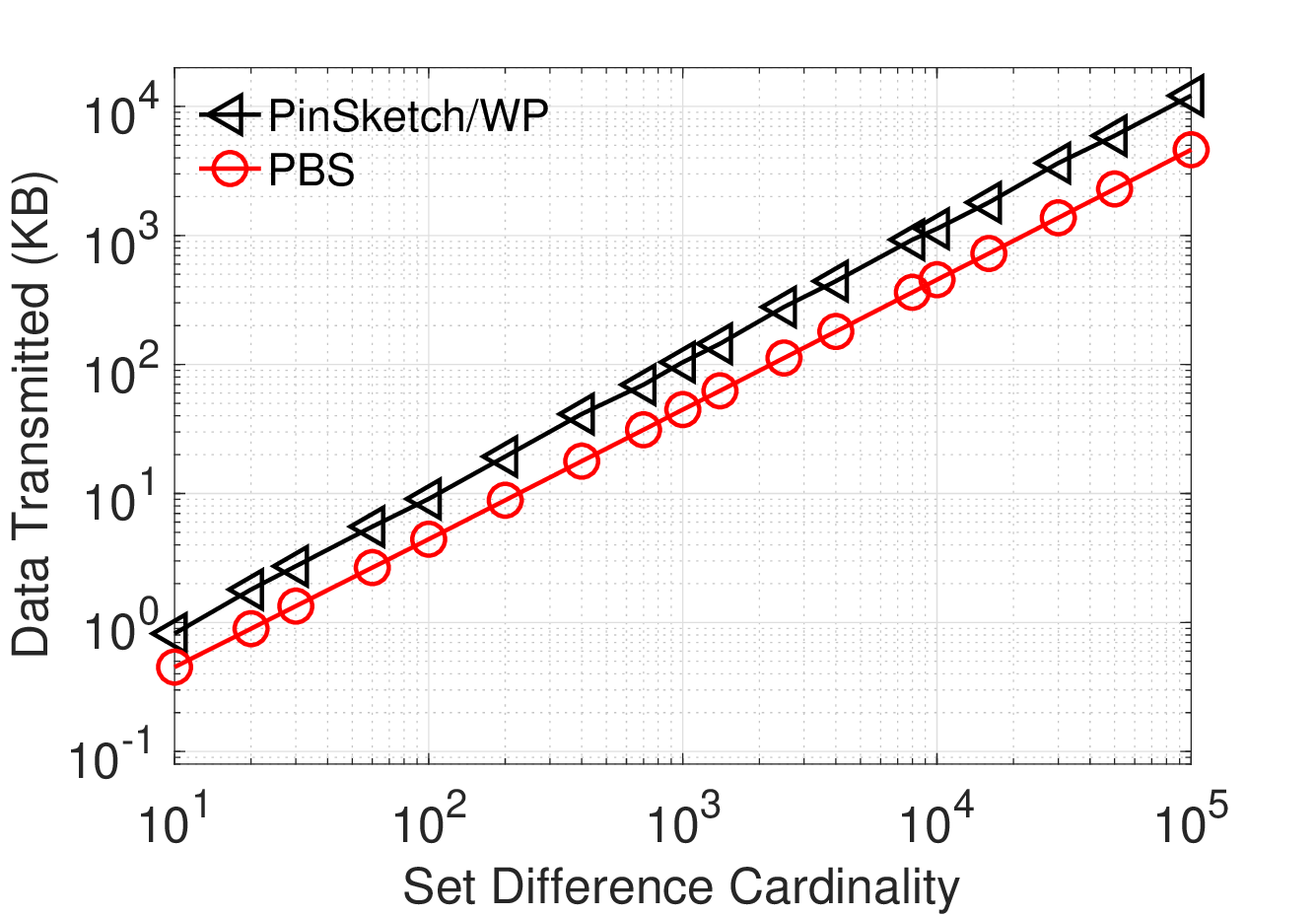}
\caption{Comparisons against PinSketch/WP, with a target success rate of 0.99 ($\log|\mathcal{U}|=$\SI{256}{bits}). }
\label{fig:vs-pinsketchwp}
\end{figure}

{\color{red}In this section, we investigate the empirical number of rounds required by PBS to correctly reconcile 
all distinct elements. The parameter settings are exactly the same as those we used in~\autoref{subsec:vs-ddigest-pinsketch}. 
The only difference is that we let PBS run as many as rounds it requires instead of only allowing it running at most 
$3$ rounds. 

\autoref{tab:completed-rounds} presents the empirical distributions of the number of rounds required by 
PBS to correctly reconcile all distinct elements, with the set difference cardinality $d$=\numlist{10;100;1000;10000;100000}. 
It is easy to verify that the average numbers of rounds are \numlist{1.20;1.81;2.04;2.09;2.18} for $d$=\numlist{10;100;1000;10000;100000} respectively.
Furthermore, in every experiment the reconciliation process took no more than 3 rounds to complete.  
Hence the $3$ probability values in every row of~\autoref{tab:completed-rounds}
add up to (probability) $1$.} 

\subsection{PBS Performances When Varying \texorpdfstring{$\delta$}{Delta}}\label{app:vs-delta}

{\color{red}
In this section, we investigate the performance of PBS under different values of 
$\delta$, where $\delta$ is the average number of distinct elements per group pair.
We have considered $\delta$ a constant throughout this paper, but in this section only, we consider it a tunable parameter.
In our experiments, we vary $\delta$ between $3$ and 
$30$.  
Like in~\autoref{subsec:vs-ddigest-pinsketch}, for each $\delta$ value, 
we set the values of the other parameters of PBS in such a way that it guarantees to correctly reconcile all 
distinct elements in no more than $r=3$ rounds with a probability of at least $p_0=0.99$.
We have experimented with different values of $d$ (the set difference cardinality).  Here we only 
present the results for $d$=\num{10000}, as other values of $d$ lead to similar conclusions. 

The experimental results, shown in~\autoref{fig:comp99-diff-avgs}, confirm our earlier claim that $\delta$ 
can serve as a knob to control the tradeoff between communication and computational overheads in PBS.
\autoref{fig:co-99-diff-avg} shows that the communication overhead of PBS generally decreases 
as $\delta$ grows, whereas \Cref{fig:et-99-diff-avgs,fig:dt-99-diff-avgs} show that both 
the encoding time and the decoding time of PBS increase as $\delta$ grows. 
}

\subsection{PBS vs. PinSketch with Partition}\label{app-sec:vs-pinsketchwp}

{\color{red}In this section, we compare again the performance of PBS and PinSketch/WP when the 
hash signature length $\log|\mathcal{U}|=$\SI{256}{bits}. 
As the implementations of all the evaluated algorithms do not support signature length higher 
than $64$ bits (one of them only supports 
32 bits), the results here are obtained through simulations with a $32$-bit universe. 
Thus, we only present the results for communication overheads. 
The results are shown in~\autoref{fig:vs-pinsketchwp}.

\autoref{fig:vs-pinsketchwp} (comparing with \autoref{fig:co-99-pswp}) clearly shows that the outperformance of 
PBS over PinSketch/WP is more significant (than that in \autoref{fig:co-99-pswp}).}

\end{document}